\documentclass[journal]{IEEEtran}

\usepackage{cite}
\usepackage{graphicx}
\usepackage{algorithm,algpseudocode}

\usepackage{amssymb}
\usepackage{mathrsfs}
\usepackage{amsmath}

\usepackage{stmaryrd}
\usepackage{pifont}
\usepackage{wasysym}

\usepackage{array}
\usepackage{url}
\usepackage{color}
\usepackage{caption}
\usepackage{multirow}
\usepackage{subcaption}
\usepackage{enumitem}
\usepackage{tabularx,adjustbox}

\algnewcommand{\LeftComment}[1]{\Statex \(\triangleright\) #1}

\makeatletter
\newcommand{\algmargin}{\the\ALG@thistlm}
\makeatother
\newlength{\whilewidth}
\algdef{SE}[parWHILE]{parWhile}{EndparWhile}[1]
{\parbox[t]{\dimexpr\linewidth-\algmargin}{%
		\hangindent\whilewidth\strut\algorithmicwhile\ #1\ \algorithmicdo\strut}}{\algorithmicend\ \algorithmicwhile}%
\algnewcommand{\parState}[1]{\State%
	\parbox[t]{\dimexpr\linewidth-\algmargin}{\strut #1\strut}}

\newtheorem{definition}{Definition}

\newtheorem{innercustomthm}{Theorem}
\newenvironment{customthm}[1]
{\renewcommand\theinnercustomthm{#1}\innercustomthm}
{\endinnercustomthm}

\begin{document}

\title{Decentralized Ride-Sharing and Vehicle-Pooling Based on Fair Cost-Sharing Mechanisms}

\author{Sid Chi-Kin Chau, Shuning Shen and Yue Zhou
\thanks{S. C.-K. Chau, S. Shen and Y. Zhou are with the Research School of Computer Science, Australian National University. (Email: sid.chau@anu.edu.au).}
\thanks{This paper appears in IEEE Trans. on Intelligent Transportation Systems.}}

\maketitle

\begin{abstract}
Ride-sharing or vehicle-pooling allows commuters to team up spontaneously for transportation cost sharing. This has become a popular trend in the emerging paradigm of sharing economy. One crucial component to support effective ride-sharing is the matching mechanism that pairs up suitable commuters. Traditionally, matching has been performed in a centralized manner, whereby an operator arranges ride-sharing according to a global objective (e.g., total cost of all commuters). However, ride-sharing is a decentralized decision-making paradigm, where commuters are self-interested and only motivated to team up based on individual payments. Particularly, it is not clear how transportation cost should be shared fairly between commuters, and what ramifications of cost-sharing are on decentralized ride-sharing. This paper sheds light on the principles of decentralized ride-sharing and vehicle-pooling mechanisms based on stable matching, such that no one would be better off to deviate from a stable matching outcome. We study various fair cost-sharing mechanisms and the induced stable matching outcomes. We compare the stable matching outcomes with a social optimal outcome (that minimizes total cost) by theoretical bounds of social optimality ratios, and show that several fair cost-sharing mechanisms can achieve high social optimality. We also corroborate our results with an empirical study of taxi sharing under fair cost-sharing mechanisms by a data analysis on New York City taxi trip dataset, and provide useful insights on effective decentralized mechanisms for practical ride-sharing and vehicle-pooling.

\end{abstract}


\section{Introduction}

There is an increasing notion of shareability in transportation systems. The popular trend of on-demand ride-hailing services (e.g., Uber, Lyft and Didi) allows commuters to arrange chauffeured vehicle services conveniently on online platforms. Certain ride-hailing platforms also provide sharing services among multiple commuters (e.g., UberPool, Lyft Line, Didi Hitch). Sharing rides is a prominent example of sharing economy, which promotes economic sharing activities in a peer-to-peer manner. Furthermore, worldwide governments are introducing policies to encourage vehicle-pooling. Private vehicles are often occupied by single passengers. Vehicle-pooling is an effective solution to improve traffic congestion, air quality and parking availability.

Despite the promising benefits, it is not clear whether commuters will be motivated to adopt ride-sharing and vehicle-pooling themselves. While there have been extensive studies \cite{S14pnas} suggesting a significant reduction in the total transportation cost by centralized ride-sharing arrangement, it is unlikely that commuters will conform to centralized arrangement without considering their individual payments. In particular, ride-sharing is a decentralized decision-making paradigm. Commuters are often self-interested and only motivated to team up with each other based on individual objectives. There is a recent trend of using social networks for arranging ride-sharing. The openness of social networks will foster a user-centric decentralized paradigm of ride-sharing and vehicle-pooling without intermediary control of third parties.
Also, many existing commercial ride-hailing service platforms are restricted to matching commuters with similar trips along the same direction. Matching non-collocated commuters is significantly more challenging. This paper sheds light on how decentralized mechanisms should be designed to support effective ride-sharing and vehicle pooling among non-collocated commuters. We aim to provide the theoretical foundation for decentralized mechanisms, as a departure from the centralized and restrictive mechanisms provided in nowadays ride-hailing service platforms. A thorough understanding of decentralized mechanisms not only benefits users, but also enables socially transparent mechanisms for our society and helps transportation companies to adopt more democratized mechanisms.

\begin{figure}[h!]
\centering
\includegraphics[width=0.5\textwidth]{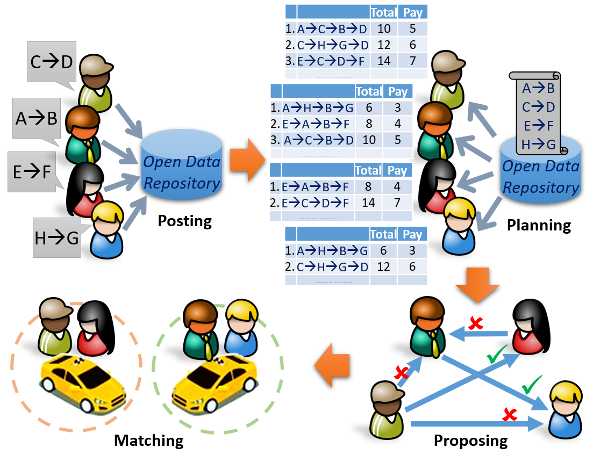} 
\caption{Decentralized ride-sharing arrangement process.} \label{fig:matching}
\end{figure}

To illustrate the concept of decentralized ride-sharing arrangement process, we provide an example\footnote{This ride-sharing arrangement process can be operated as a standalone services on top of any ride-hailing platforms (like Uber or taxis).} in Fig.~\ref{fig:matching}. First, the commuters will post their trip information and time constraints on an open data repository (e.g.,  a social network or an open ledger like blockchain). Then, the commuters will identify potential ride-sharing  partners and plan the possible shared rides with transportation costs. The commuters will also compute their individual payments by splitting the transportation costs in a certain fair manner. Next, the commuters will propose to potential ride-sharing partners according to the ranking order of individual payments. By a proper matching mechanism, they will reach a mutual agreement, such that no better rides can be arranged otherwise. Note that this process is not dictated by a centralized operator. Each commuter is free to accept or reject any ride-sharing proposals.

\medskip

In particular, we highlight three key elements in decentralized ride-sharing arrangement:

\subsection{Fair Cost-Sharing Mechanisms}

Central to decentralized ride-sharing arrangement is a commonly agreed {\em cost-sharing mechanism} for splitting the transportation costs among the parties of ride-sharing. The choice of cost-sharing mechanisms should take into consideration of {\em fair} contribution of each party,  which provides a rationale on how to split the cost of a sharable ride in a fair manner.  In particular, commuters may not share the same destinations or sources. There are a variety of possible fair cost-sharing mechanisms. For example, one simple cost-sharing mechanism is to split the the transportation cost equally between two parties. Another way is to split proportionally according to the original costs of standalone rides. Also, one may consider to split in a way to induce equal savings from standalone rides. Note that the choices of cost-sharing mechanisms will affect individual commuters' preferential orders of possible ride-sharing options and the outcome of an agreement. Given multiple choices of cost-sharing mechanisms, it is important to understand their ramifications on ride-sharing agreement.

\subsection{Stable Matching}

Based on individual commuters' preferential orders of possible ride-sharing options, a decentralized matching mechanism is needed to arrange ride-sharing. In practice, commuters usually form a coalition as a pair to share a ride, as this reduces the complexity of reaching an agreement. Matching mechanisms have been studied in various applications, such as college admissions and dating \cite{stablematchbook}. One useful concept is {\em stable matching}, which is particularly desirable in decentralized decision-making mechanisms, because no participants would be better off to deviate from a stable matching outcome. Hence, stable matching captures the likely outcome of an agreement in a decentralized matching process. But different cost-sharing mechanisms will induce different stable matching outcomes. In this paper, we compare stable matching outcomes under different fair cost-sharing mechanisms in ride-sharing arrangement.

\subsection{Social Optimality}

To compare different cost-sharing mechanisms for ride-sharing arrangement, a natural approach is to benchmark against a {\em social optimal} outcome that minimizes the total transportation cost of all commuters. Decentralized ride-sharing arrangement will not reach a social optimal outcome, because everyone is self-interested to minimize his individual cost, rather than the total cost. However, a good cost-sharing mechanism should achieve high social optimality. We measure the ratio between the cost of a stable matching outcome over the one of a social optimal outcome. In this paper, we present theoretical bounds on the {\em social optimality ratios} for various fair cost-sharing mechanisms, which then show high social optimality in these cost-sharing mechanisms. To corroborate our theoretical study, we also present a data analysis on practical taxi sharing in New York City. We compare the empirical social optimality of various fair cost-sharing mechanisms used in taxi sharing with NYC taxi trip dataset \cite{nycdb}.

\medskip

{\em Outline:} This paper presents an extensive study of how decentralized mechanisms can support effective ride-sharing. We first present a brief survey on the related literature in Sec.~\ref{sec:related}. The model and notations of cost-sharing mechanisms and stable matching are formulated in Secs.~\ref{sec:model}-\ref{sec:mechanism}. We compare the social optimality of various fair cost-sharing mechanisms by theoretical bounds in Sec.~\ref{sec:PoA}. We also present a modified stable matching algorithm for finding a stable ride-sharing assignment in Sec.~\ref{sec:matching}. To corroborate our theoretical study, we present an empirical data analysis in Sec.~\ref{sec:empirical}.

\section{Related Work} \label{sec:related}

Ride-sharing research belongs to a large body of literature. There are various paradigms of ride-sharing among commuters, or between drivers and passengers. For example, see a survey in \cite{NR2016ridesharing}. One of the critical components in ride-sharing is the matching process that pairs commuters to share a ride, or finding suitable drivers for the requested passengers  \cite{ZZ19matching, P15matching, W17matching,agatz2020stableride,wang2017matching,RC2019matching,peng2020stableride}. 
In particular, the studies in \cite{S14pnas, A17pnas} investigated how ride-sharing and vehicle pooling can reduce transportation delay. These papers generally assume that ride-sharing and vehicle pooling is arranged by centralized entities to achieve the desirable benefits. There is no consideration of the self-interested nature of commuters who will not always follow centralized arrangement. On the other hand, there are recent papers considering stable matching in ride-sharing \cite{ZZ19matching, P15matching, W17matching}. However, the motivation of stable matching in these papers is related to arbitrary passengers' or drivers' preferences about each other, which is not necessarily related to cost-sharing of transportation costs. \cite{agatz2020stableride,wang2017matching,RC2019matching,peng2020stableride} consider stable matching between drivers and passengers, but do not consider sharing transportation cost among the passengers. Also, these papers did not compare stable matching with social optimal ride-sharing arrangements in the context of sharing transportation cost among commuters.
%Ride-sharing belongs to the general topic of transportation scheduling. Scheduling algorithms for ride-sharing have been investigated in other studies \cite{Z19rideshare, M15rideshare}, where a scheduler optimizes a ride to pick up and drop off multiple passengers at different locations. While optimization of ride is outside the scope of the paper, the scheduling algorithms can be incorporated in the matching algorithm to arrange the best shared rides among a given set of commuters.

Our study of fair cost-sharing mechanisms for ride-sharing belongs to the broad topic of network cost-sharing and coalition formation problems. A study related to our results is the strong price of anarchy for stable matching \cite{ADN09}. Our ride-sharing matching problem is a subclass of coalition formation games \cite{AB12} that allows arbitrary coalitions with at most two participants per coalition. However, our results are based on social optimality ratio, which is different than the previous studies. The results of social optimality ratio in this work can be derived in a general setting in \cite{CE17sharing, CE20sharing, chau19p2penergy}. But we simplify the proofs by considering a coalition as pairs in the context of ride-sharing. Furthermore, we corroborate our theoretical study by an empirical data analysis of practical taxi sharing in New York City.

The New York City taxi trip dataset is a large publicly available dataset \cite{nycdb}, which provides detailed records of pick-up and drop-off locations and times in New York City. The dataset can enable a wide range of empirical studies of taxi service strategy optimization \cite{cmtseng2019etaxi}. There have been a number of studies about taxi sharing in the literature based on New York City taxi trip dataset \cite{S14pnas, A17pnas}. But none of the previous studies considered decentralized stable matching for taxi sharing using New York City taxi trip dataset.

%%%%%%%%%%%%%%%%%%%%%%%%%%%%%%%%%%%%%%%
\begin{table}[t!]
    \centering\caption{\label{tbl:symbs}Table of key symbols and notations.}
    \scalebox{1}{\begin{tabularx}{\linewidth}{@{}r@{\ }|@{}X@{}}
    \hline    \hline
    Symbol & \ Definition \\
    \hline
   ${\cal G}_{\rm road}= ({\cal V}_{\rm road}, {\cal E}_{\rm road})$  & $\circ$ Graph representing road network  \\
   ${\cal N}$  & -Set of commuters participating in ride-sharing \\
   $(v^{\rm s}_i, v^{\rm d}_i, t^{\rm s}_i, t^{\rm d}_i)$  &$\circ$ Source and destination locations, and earliest departure and latest arrival times of  $i \in {\cal N}$\\ 
   ${\cal R}_{i, j}$ &$\circ$ Set of sharable rides for $i, j \in {\cal N}$ satisfying feasibility constraints \\
   $r^{\rm self}_{i}$ &$\circ$ Standalone ride of $i \in {\cal N}$ not shared with others\\      
   ${\cal G}_{\rm match} = ({\cal N}, {\cal E}_{\rm match})$  &$\circ$ Matching graph, where $(i, j) \in {\cal E}_{\rm match}$ has non-empty ${\cal R}_{i, j}$ and $(i, i) \in {\cal E}_{\rm match}$ for all $i \in {\cal N}$\\
   $r^{\min}_{i, j}$ &$\circ$ Minimum cost sharable ride in ${\cal R}_{i, j}$\\
   $c_{i, j} \triangleq c( r^{\min}_{i, j} )$ &$\circ$ Cost of shared ride $r^{\min}_{i, j}$ for $i, j \in {\cal N}$\\
   $c_{i,i} \triangleq c( r^{\rm self}_{i} )$ &$\circ$ Cost of standalone ride $r^{\rm self}_{i}$ for $i \in {\cal N}$\\
   $p_i(r)$ &$\circ$Payment from $i$ for shared ride $r$ based on a cost-sharing mechanism\\   
   $u_i(r) \triangleq c(r^{\rm self}_{i}) - p_i(r)$ &$\circ$ Utility of $i$ for of shared ride $r$\\   
   ${\cal M}$ &$\circ$ Feasible ride-sharing assignment, a subset of ${\cal E}_{\rm match}$ satisfying the feasibility properties\\   
   $c({\cal M}) \triangleq \sum_{(i,j) \in {\cal M}} c_{i,j}$ &$\circ$ Social cost of ride-sharing assignment ${\cal M}$\\   
   $\frac{c(\hat{\cal M})}{c({\cal M}^\ast)}$ &$\circ$ Social optimality ratio between a stable ride-sharing assignment ($\hat{\cal M}$) and a social optimal ride-sharing assignment (${\cal M}^\ast$) \\
    \hline    \hline
    \end{tabularx}}
\end{table}
%%%%%%%%%%%%%%%%%%%%%%%%%%%%%%%%%%%%%%%

\section{Model and Notations} \label{sec:model}

This section presents a general model of ride-sharing by matching commuters for sharing hired vehicles\footnote{In practice, the matching process may be carried out on an online platform automated by computer agents representing the users.}. Table~\ref{tbl:symbs} lists some key notations. Consider a road network represented by a directed graph ${\cal G}_{\rm road}= ({\cal V}_{\rm road}, {\cal E}_{\rm road})$, where ${\cal V}_{\rm road}$ is a set of road junctions and ${\cal E}_{\rm road}$ is a set of road segments. There is a set of commuters ${\cal N}$. Each commuter $i \in {\cal N}$ is associated with a tuple of parameters $(v^{\rm s}_i, v^{\rm d}_i, t^{\rm s}_i, t^{\rm d}_i)$, where $v^{\rm s}_i \in {\cal V}_{\rm road}$ is the source location, $v^{\rm d}_i \in {\cal V}_{\rm road}$ is the destination location, $t^{\rm s}_i$ is the earliest departure time, and $t^{\rm d}_i$ is the latest arrival time. Our goal is to pair up the commuters for potential ride-sharing.

\subsection{Sharable Rides}

Given road network ${\cal G}_{\rm road}$, a {\em ride} $r$ is defined by a sequence of locations $(v^r_1, ...,v^r_l, ..., v^r_m)$, where each $v^r_j \in {\cal V}_{\rm road}$,  and a sequence of arrival times $(t^r_1, ..., t^r_l, ..., t^r_m)$, where $t^r_l$ is the arrival time at location $v^r_l$. For each $v \in  \{v^r_1, ..., v^r_m\}$, we denote $t^r(v)$ as the arrival time of ride $r$ at location $v$.

\medskip

\begin{definition}({\bf Sharable Ride})
A ride $r$ is {\em sharable} by a pair $i, j \in {\cal N}$, if the following feasibility constraints are satisfied:
\begin{enumerate}

\item ({\em Location Constraint}): $( v^{\rm s}_i, v^{\rm d}_i, v^{\rm s}_j, v^{\rm d}_j )$ are in the sequence of locations of ride $r$. Namely, $v^{\rm s}_i, v^{\rm d}_i, v^{\rm s}_j, v^{\rm d}_j \in \{v^r_1, ..., v^r_m\}$. 

\item ({\em Temporal Constraint for $i$}): $t^{\rm s}_i \le t^r(v^{\rm s}_i) < t^r(v^{\rm d}_i) \le t^{\rm d}_i$.

\item ({\em Temporal Constraint for $j$}): $t^{\rm s}_j \le t^r(v^{\rm s}_j) < t^r(v^{\rm d}_j) \le t^{\rm d}_j$.

\end{enumerate}

Given $(i, j)$, there are two types of sharable rides:
\begin{enumerate}

\item {\em Hitchhiking Ride}:
A ride $r$ is called an $(i; j)$-hitchhiking ride, if $t^r(v^{\rm s}_j) < t^r(v^{\rm s}_i) < t^r(v^{\rm d}_i) < t^r(v^{\rm d}_j)$.

\item {\em Combined Ride}:
A ride $r$ is called an $(i; j)$-combined ride, if $t^r(v^{\rm s}_i) < t^r(v^{\rm s}_j) < t^r(v^{\rm d}_i) < t^r(v^{\rm d}_j)$.

\end{enumerate}
A sharable ride for $(i, j)$ can either be $(i; j)$-hitchhiking, $(j; i)$-hitchhiking, $(i; j)$-combined, or $(j; i)$-combined. For example, in Fig.~\ref{fig:example} (b), $(i, k)$ shares $(i; k)$-hitchhiking ride. In Fig.~\ref{fig:example} (c), $(i, j)$ shares $(i; j)$-combined ride.
\end{definition}

In this paper, we consider the matching of pairs of commuters, who have declared their requests in advance, barring cancellation. Also, we consider the cost of transportation as the primary factor to ride-sharing decisions. However,  the model can be extended by incorporating additional constraints in the matching process.
Given a ride $r$, let $c(r)$ be the associated transportation cost, which will be the fare of a taxi or hired vehicle. Let ${\cal R}_{i, j}$ be the set of sharable rides in road network ${\cal G}_{\rm road}$ for a pair of distinct commuters $(i, j)$, and $r^{\min}_{i, j} \triangleq \arg\min_{r \in {\cal R}_{i, j}} c(r)$ be the {\em minimum cost sharable ride} in ${\cal R}_{i, j}$, which is the ride of minimum cost among all sharable rides between $i, j$. Let $r^{\rm self}_{i}$ be the {\em standalone ride} for commuter $i$ if $i$ does not share with another commuter. Note that $c(r^{\min}_{i, j}) \ge \max\{ c(r^{\rm self}_{i}), c(r^{\rm self}_{j}) \}$. Otherwise, the commuter ($i$ or $j$) can always choose another lower cost standalone ride.

\subsection{Matching for Ride-Sharing} 

Matching for ride-sharing can be attained by an undirected matching graph ${\cal G}_{\rm match} = ({\cal N}, {\cal E}_{\rm match})$ where ${\cal E}_{\rm match} \triangleq \{ (i, j) : {\cal R}_{i, j} \ne \varnothing \} \cup \{ (i, i) : i \in {\cal N}\}$. Namely, ${\cal E}_{\rm match}$ includes two types of edges: (1) $(i, j)$ represents a sharable ride with a pair of distinct commuters, and (2) $(i, i)$ represents a standalone ride. Let $c_{i, j} \triangleq c( r^{\min}_{i, j} )$ and $c_{i,i} \triangleq c( r^{\rm self}_{i} )$ be the edge costs of edges  $(i, j)$ and $(i, i)$, respectively.

\medskip

\begin{definition}({\bf Feasible Ride-Sharing})
Given matching graph ${\cal G}_{\rm match}$, we define a {\em feasible ride-sharing assignment} as a subset of edges ${\cal M} \subseteq {\cal E}_{\rm match}$ satisfying the following feasibility properties:
\begin{enumerate}

\item
Every $i \in {\cal N}$ is covered by an edge in ${\cal M}$. Namely, there exists $(i, j) \in {\cal M}$ or $(i, i) \in {\cal M}$ for every $i \in {\cal N}$.

\item
No pair of edges in ${\cal M}$ share any nodes. Namely, there do not exist $(i, j) \in {\cal M}$ and $(i, k) \in {\cal M}\backslash\{(i, j)\}$ for $i \in {\cal N}$.

\end{enumerate}
\end{definition}

\begin{definition}({\bf Socially Optimal Ride-Sharing})
Given a feasible ride-sharing assignment ${\cal M}$, let the social cost be $c({\cal M}) \triangleq \sum_{(i,j) \in {\cal M}} c_{i,j}$. A feasible ride-sharing assignment is called a {\em social optimum} (denoted by ${\cal M}^\ast$), if it minimizes the total transportation cost:
\begin{align}
& \min_{{\cal M}: {\cal M} \mbox{\ is feasible}} c({\cal M}) \notag 
\end{align}

A ride-sharing assignment ${\cal M}$ can also be equivalently represented by a binary vector $x = (x_{i,j})_{(i,j) \in {\cal E}_{\rm match}}$, where each binary variable $x_{i,j} \in \{0, 1\}$ indicates whether the pair of commuters $(i, j)$ will share a ride. Note that finding a social optimal ride-sharing assignment is equivalent to solving the following minimum weight edge covering problem:
\begin{align}
& \min_{x} \sum_{(i,j) \in {\cal E}_{\rm match}} c_{i,j} x_{i,j}  \notag \\
\mbox{subject to\ }  & \sum_{j\in {\cal N}} x_{i,j}  \ge 1, \mbox{\ for all\ } i \in {\cal N} \label{cons:cover1} \\
& x_{i,j}  \in \{0, 1\}, \mbox{\ for all\ } (i,j) \in {\cal E}_{\rm match} \label{cons:cover2}
\end{align}
Constraints~(\ref{cons:cover1})-(\ref{cons:cover2}) ensure the feasibility of ride-sharing assignment.
\end{definition}

\medskip

\begin{figure}[t!]
\includegraphics[width=0.5\textwidth]{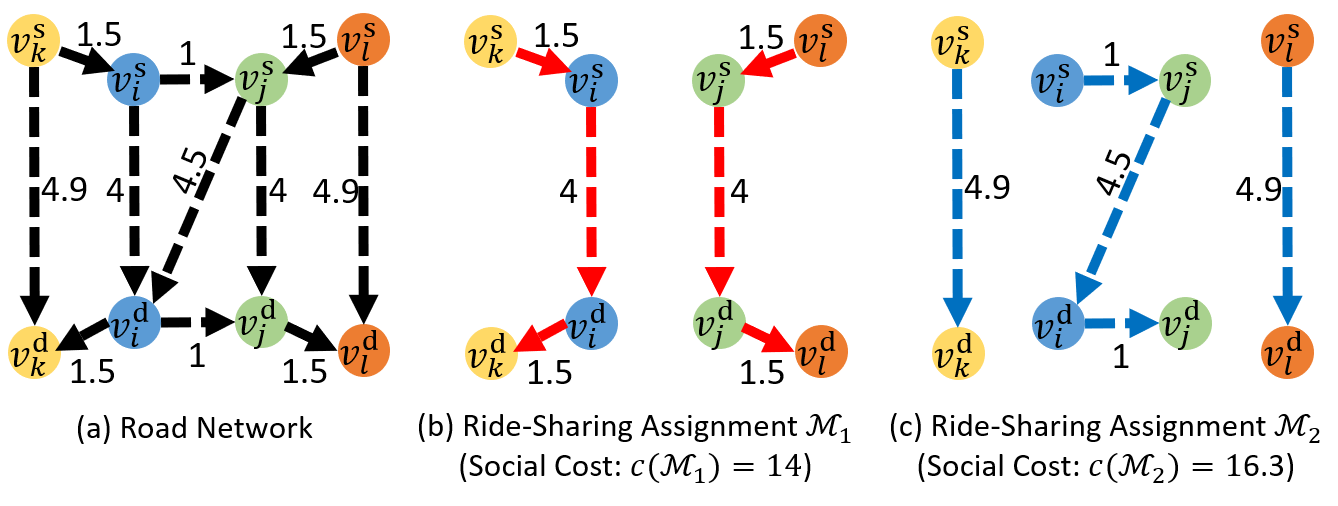} 
\caption{An example of ride-sharing assignments. ${\cal M}_1 = \{(i, k), (j, l) \}$ is social optimum. ${\cal M}_1$ is stable under egalitarian and proportional cost-sharing mechanisms (to be defined in Sec.~\ref{sec:mechanism}), whereas ${\cal M}_2 = \{(i, j), (k), (l) \}$ is stable under equal and segment-based cost-sharing mechanisms.} \label{fig:example}
\end{figure}

{\bf Example:} We consider an example of road network in Fig.~\ref{fig:example} (a). There are four commuters $\{i, j, k, l\}$. For commuter $i$, his sources location is denoted by $v^{\rm s}_i$ and destination location by $v^{\rm s}_j$. The number on each edge represents the transportation cost of the respective segment. In this example, the social optimal ride-sharing assignment is ${\cal M}_1 = \{(i, k), (j, l) \}$ with social cost $c({\cal M}_1) = 14$, as illustrated in Fig.~\ref{fig:example} (b).

\section{Decentralized Ride-Sharing Mechanisms} \label{sec:mechanism}

Note that ride-sharing is a decentralized decision-making paradigm. It is unlikely that every commuter will follow a ride-sharing assignment according to the social cost. In reality, commuters are self-interested and only motivated to team up with each other based on individual savings. It is natural to consider how individual commuter will decide in arranging ride-sharing among themselves.

\subsection{Fair Cost-Sharing Mechanisms}

Consider a sharable ride $r$ for a pair of commuters $(i, j)$. There are many ways to split to cost $c(r)$ among $(i, j)$ in a fair manner. A cost-sharing mechanism is defined by payment function $p_i(\cdot)$, which is the payment by commuter $i$ for ride $r$. A {\em budget-balanced} cost-sharing mechanism requires $p_i(r) + p_j(r) = c(r)$. We also let the utility (i.e., saving) of commuter $i$ for ride $r$ be $u_i(r) = c(r^{\rm self}_{i}) - p_i(r)$, where $r^{\rm self}_{i}$ is the standalone ride for commuter $i$. 

A fair cost-sharing mechanism should provide a rationale on how to split the cost of a sharable ride in a fair manner. There are several fair and budget-balanced cost-sharing mechanisms (denoted by different superscripts in $p_i(\cdot)$) as follows:
\begin{enumerate}

\item ({\bf Equal Cost-Sharing}): 
Each commuter should split $c(r)$ equally.
\begin{equation}
p^{\rm eq}_i(r) = \frac{c(r)}{2}
\end{equation}
Note that equal cost-sharing mechanism may produce negative utility ($u^{\rm eq}_i(r) < 0$).

\item ({\bf Egalitarian Cost-Sharing}): 
Each commuter should split $c(r)$ in a way to attain the same utility.
\begin{equation}
p^{\rm ega}_i(r)= \frac{c(r) + c(r^{\rm self}_{i}) - c(r^{\rm self}_{j})}{2}
\end{equation}
Namely, $u^{\rm ega}_i(r) = u^{\rm ega}_j(r)  = \frac{c(r^{\rm self}_{i}) + c(r^{\rm self}_{j}) - c(r)}{2}$

\item ({\bf Proportional Cost-Sharing}): 
Each commuter should contribute proportionally to the cost of standalone ride.
\begin{equation}
p^{\rm pp}_i(r) = \frac{c(r^{\rm self}_{i})\cdot c(r)}{c(r^{\rm self}_{i}) + c(r^{\rm self}_{j})}
\end{equation}

\item ({\bf Segment-based Cost-Sharing}): 
Each commuter should only contribute to the participated segments of ride. Let the transportation cost of the segment from $v_1$ to $v_2$ in ride $r$ be $c_r(v_1, v_2)$.
\begin{itemize}

\item If $r$ is an $(i; j)$-hitchhiking ride, then
\begin{align}
p^{\rm sb}_i(r) &\ = \frac{c_r(v^{\rm s}_i, v^{\rm d}_i)}{2}  \\
p^{\rm sb}_j(r) &\ =  c_r(v^{\rm s}_j, v^{\rm s}_i) + \frac{c_r(v^{\rm s}_i, v^{\rm d}_i)}{2} + c_r(v^{\rm d}_i, v^{\rm d}_j)
\end{align}

\item If $r$ is an $(i; j)$-combined ride, then 
\begin{align}
p^{\rm sb}_i(r) &\ =  c_r(v^{\rm s}_i, v^{\rm s}_j) + \frac{c_r(v^{\rm s}_j, v^{\rm d}_i) }{2} \\
p^{\rm sb}_j(r) &\ = \frac{c_r(v^{\rm s}_j, v^{\rm d}_i) }{2} + c_r(v^{\rm d}_i, v^{\rm d}_j) 
\end{align}

\end{itemize}

\end{enumerate}

Each of these fair cost-sharing mechanisms captures a notion of fair contribution from the involved commuters. Different fair cost-sharing mechanisms will induce different outcomes in decentralized ride-sharing arrangement. We next investigate the impacts of these fair cost-sharing mechanisms. In the following, we consider the sharable ride for each pair pf commuters $(i, j)$ to be the minimum cost sharable ride $r^{\min}_{i, j}$.
 
\subsection{Stable Ride-Sharing Assignment}
 
Given a cost-sharing mechanism, any pair of commuters may join to share a ride based on individual payments. Every commuter aims to minimize his individual payment. We consider any unilateral switch that allows any pair of commuters abandon their current rides to create another ride.  As a consequence, a stable assignment is likely to emerge, such that no one would be better off to deviate from the current assignment. 

\medskip

\begin{definition}({\bf Stable Ride-Sharing})
Given payment function $p_{i}(\cdot)$, a pair of commuters $(i,j)$ is called a {\em blocking pair} with respect to ride-sharing assignment ${\cal M}$ if %$G \notin {\cal M}$, and 
$i$ and $j$ can {\it strictly} reduce their payments when they form a pair $(i,j)$ to share a ride instead of the respective rides in ${\cal M}$. A feasible ride-sharing assignment $\hat{\cal M}$ is called {\em stable ride-sharing assignment}, if there exists no blocking pair with respect to $\hat{\cal M}$. Stable ride-sharing assignment is based on the concept of stable coalition in cooperative game theory \cite{CE17sharing, CE20sharing}.
\end{definition}

A stable ride-sharing assignment is equivalent to a stable matching outcome (with the inclusion of possibly singleton groups for standalone commuters). Note that a stable ride-sharing assignment  is also a strong Nash equilibrium in game theory \cite{CE17sharing}. A strong Nash equilibrium is a Nash equilibrium, in which no group of players can cooperatively deviate in an allowable way that benefits all of its members, whereas a Nash equilibrium only allows a player to make a unilateral action. Some previous papers  (e.g., \cite{agatz2020stableride}) studied Nash equilibrium the matching in ride-sharing, but not strong Nash equilibrium.

\medskip

\begin{definition}({\bf Social Optimality Ratio})
Define the social optimality ratio as the ratio between the cost of a stable ride-sharing assignment ($\hat{\cal M}$) and that of a social optimal ride-sharing assignment (${\cal M}^\ast$) for a particular instance of ride-sharing problem by
\begin{equation}
\frac{c(\hat{\cal M})}{c({\cal M}^\ast)}
\end{equation}
If the social optimality ratio is small for a particular cost-sharing mechanism, then such a cost-sharing mechanism can achieve high social optimality by inducing a stable ride-sharing assignment close to a social optimum.
\end{definition}

\medskip

\begin{table}[th!]
\begin{center}

\begin{tabular}{c|cccc}
\hline \hline 
& eq & ega & pp & sb \\
\hline 
$p_i(r^{\min}_{i, j})$ & 3.25 & 3.25 & 3.25 & 3.25 \\
$p_j(r^{\min}_{i, j})$ & 3.25 & 3.25 & 3.25 & 3.25 \\
$p_i(r^{\min}_{i, k})$ & 3.5 & 3 & 3.11 & 2 \\
$p_k(r^{\min}_{i, k})$ & 3.5 & 4 & 3.89 & 5 \\
$p_j(r^{\min}_{j, l})$ & 3.5 & 3 & 3.11 & 2 \\
$p_l(r^{\min}_{j, l})$ & 3.5 & 4 & 3.89 & 5 \\
\hline
stable ride-sharing & ${\cal M}_2$ & ${\cal M}_1$ & ${\cal M}_1$ & ${\cal M}_2$ \\
\hline \hline
\end{tabular}

\medskip

\begin{tabular}{c|c|c|c}
\hline \hline 
 $c(r^{\rm self}_i)$ & $c(r^{\rm self}_j)$ & $c(r^{\rm self}_k)$ & $c(r^{\rm self}_l)$ \\
 \hline
4 & 4 & 4.9 & 4.9 \\
\hline \hline
\end{tabular}

\caption{Individual payments based on different cost-sharing mechanisms (eq, ega, pp, sb) and the costs of standalone rides for the example in Fig.~\ref{fig:example}.} \label{tab:example}
\end{center}
\end{table}

{\bf Example:} We consider the road network in Fig.~\ref{fig:example} (a). Table~\ref{tab:example} shows the individual payments based on  equal (eq), egalitarian (ega), proportional (pp), and segment-based (sb)  cost-sharing mechanisms for commuters $\{i ,j, k, l \}$, and the costs of standalone rides. The stable ride-sharing assignments for different cost-sharing mechanisms are derived as follows:
\begin{enumerate}

\item ({\em Equal Cost-Sharing}): We obtain 
\begin{align}
p_i^{\rm eq}(r^{\min}_{i, j}) <\ & p_i^{\rm eq}(r^{\min}_{i, k}) < c(r^{\rm self}_i) \\
p_j^{\rm eq}(r^{\min}_{i, j}) <\ & p_j^{\rm eq}(r^{\min}_{j, l}) < c(r^{\rm self}_j)
\end{align}
Hence, $(i, j)$ will be motivated to share a ride instead of other options. Then, the stable ride-sharing assignment is ${\cal M}_2 = \{ (i, j), (k), (l)\}$ and the social optimality ratio is 1.16.

\item ({\em Egalitarian Cost-Sharing}): We obtain 
\begin{align}
p_i^{\rm ega}(r^{\min}_{i, k}) <\ & p_i^{\rm ega}(r^{\min}_{i, j}) < c(r^{\rm self}_i) \\
p_k^{\rm ega}(r^{\min}_{i, k}) <\ & c(r^{\rm self}_k) \\
p_j^{\rm ega}(r^{\min}_{j, l}) <\ & p_j^{\rm ega}(r^{\min}_{i, j}) < c(r^{\rm self}_j) \\
p_l^{\rm ega}(r^{\min}_{j, l}) <\ & c(r^{\rm self}_l)
\end{align}
Hence, $(i, k)$ and $(j, l)$ will be motivated to share rides respectively. The stable ride-sharing assignment is ${\cal M}_1 = \{ (i, k), (j, l)\}$ and the social optimality ratio is 1.

\item ({\em Proportional Cost-Sharing}): Similar to egalitarian cost-sharing, the stable ride-sharing assignment is ${\cal M}_1 = \{ (i, k), (j, l)\}$ and the social optimality ratio is 1.

\item ({\em Segment-based Cost-Sharing}): We obtain 
\begin{align}
c(r^{\rm self}_k)  <\ & p_k^{\rm sb}(r^{\min}_{i, k}) \\
c(r^{\rm self}_l)  <\ & p_l^{\rm sb}(r^{\min}_{j, l}) \\
p_i^{\rm sb}(r^{\min}_{i, k}) <\ & p_i^{\rm sb}(r^{\min}_{i, j}) < c(r^{\rm self}_i)  \\
p_j^{\rm sb}(r^{\min}_{j, l}) <\ & p_j^{\rm sb}(r^{\min}_{i, j}) < c(r^{\rm self}_j) 
\end{align}
Hence, $k$ and $l$ will be motivated to remain standalone rides, and then $(i, j)$ will share a ride. The stable ride-sharing assignment is ${\cal M}_2 = \{ (i, j), (k), (l)\}$ and the social optimality ratio is 1.16.

\end{enumerate}

{\bf Remark:} 
In the preceding example, egalitarian and proportional cost-sharing mechanisms can induce a social optimal stable ride-sharing assignment. But in general, different stable ride-sharing assignments will be induced by different cost-sharing mechanisms. It is crucial to understand the social optimality ratios of different cost-sharing mechanisms. In Sec.~\ref{sec:PoA}, we provide theoretical upper bounds on the social optimality ratios, which shows that several fair cost-sharing mechanisms can achieve high social optimality.

\section{Bounds on Social Optimality Ratio} \label{sec:PoA}

Given any feasible ride-sharing assignment ${\cal M} \subseteq {\cal E}_{\rm match}$, let the set of commuters present in ${\cal M}$ be ${\cal N}({\cal M}) \triangleq \{ i \in {\cal N} : (i,j) \in {\cal M} \}$. For $i \in {\cal N}({\cal M})$, let the payment of $i$ with respect to ${\cal M}$ be $p_i({\cal M}) = p_i(r^{\min}_{i, j})$, where $(i,j) \in {\cal M}$. In the following, we provide general theories to upper bound the social optimality ratio ($\frac{c(\hat{\cal M})}{c({\cal M}^\ast)}$) between the cost of a stable ride-sharing assignment ($\hat{\cal M}$) and that of a social optimal ride-sharing assignment (${\cal M}^\ast$) for any instances of problem. In Sec.~\ref{sec:empirical}, we will corroborate our theoretical results with an empirical data analysis on taxi sharing in New York City.

\medskip

\begin{customthm}{1} \label{thm:eq-er}
For equal cost-sharing mechanism, let $\hat{\cal M}^{\rm eq}$ be a stable ride-sharing assignment. We show that the social optimality ratio is upper bounded by $\frac{c(\hat{\cal M}^{\rm eq})}{c({\cal M}^\ast)} \le \frac{3}{2}$.
%\begin{equation}
%\frac{3}{2} \ge \frac{c(\hat{\cal M}^{\rm eq})}{c({\cal M}^\ast)}
%\end{equation}
\end{customthm}

\begin{IEEEproof}
First, we assume that $\hat{\cal M}^{\rm eq} \ne {\cal M}^\ast$. Otherwise, $\frac{3}{2} \ge 1 = \frac{c(\hat{\cal M}^{\rm eq})}{c({\cal M}^\ast)}$. 
Suppose $(i, j) \in \hat{\cal M}^{\rm eq} \backslash {\cal M}^\ast$. Then there must exist $(i, k)$ and $(j, l)$, such that $(i, k), (j, l) \in {\cal M}^\ast \backslash \hat{\cal M}^{\rm eq}$, because all commuters must belong to some sharable rides in $\hat{\cal M}^{\rm eq}$ and ${\cal M}^\ast$, and both $\hat{\cal M}^{\rm eq}$ and ${\cal M}^\ast$ are feasible.

We assume that $i \ne k$ and $j \ne l$. Note that the cases of $i = k$ or $j = l$ can be proven straightforwardly. Recall $c_{i, j} \triangleq c( r^{\min}_{i, j} )$ and $c_{i,i} \triangleq c( r^{\rm self}_{i} )$. Since $(i, k), (j, l) \in {\cal M}^\ast$, we obtain 
\begin{equation}
 p^{\rm eq}_i({\cal M}^\ast) + p^{\rm eq}_j({\cal M}^\ast) + p^{\rm eq}_k({\cal M}^\ast) + p^{\rm eq}_l({\cal M}^\ast) = c_{i, k} + c_{j, l}
\end{equation}

On the other hand, since $\hat{\cal M}^{\rm eq}$ is a stable ride-sharing assignment, we obtain
\begin{align}
 p^{\rm eq}_i(\hat{\cal M}^{\rm eq}) + p^{\rm eq}_j(\hat{\cal M}^{\rm eq}) = c_{i, j},  \\
 p^{\rm eq}_k(\hat{\cal M}^{\rm eq}) \le c_k, \quad p^{\rm eq}_l(\hat{\cal M}^{\rm eq}) \le c_l
\end{align}
Hence, it follows that	
\begin{equation}
 p^{\rm eq}_i(\hat{\cal M}^{\rm eq}) + p^{\rm eq}_j(\hat{\cal M}^{\rm eq}) + p^{\rm eq}_k(\hat{\cal M}^{\rm eq}) + p^{\rm eq}_l(\hat{\cal M}^{\rm eq}) \le c_{i, j} + c_{k} + c_{l}
\end{equation}
Because $\hat{\cal M}^{\rm eq}$ is a stable ride-sharing assignment and noting that $i \ne k$ and $j \ne l$, we obtain
\begin{align}
 \tfrac{c_{i,j}}{2} =   p^{\rm eq}_i(\hat{\cal M}^{\rm eq}) \le  p^{\rm eq}_i({\cal M}^\ast) = \tfrac{c_{i,k}}{2}  &\ \Rightarrow \ c_{i,j} \le c_{i,k} \\
 \tfrac{c_{i,j}}{2} =   p^{\rm eq}_j(\hat{\cal M}^{\rm eq}) \le  p^{\rm eq}_j({\cal M}^\ast) = \tfrac{c_{j,l}}{2}  &\ \Rightarrow \ c_{i,j} \le c_{j,l} 
\end{align}

Together, by noting that $c_{i, k} \ge \max\{c_i, c_k\}$ and $c_{j, l} \ge \max\{c_j, c_l\}$, we obtain
\begin{align}
 &  3 \big(p^{\rm eq}_i({\cal M}^\ast) + p^{\rm eq}_j({\cal M}^\ast) + p^{\rm eq}_k({\cal M}^\ast) + p^{\rm eq}_l({\cal M}^\ast) \big) \notag \\
=\ & 3(c_{i, k} + c_{j, l}) \\
\ge\ & c_{i, k} + c_{j, l} + c_k + c_l + c_k + c_l\\
\ge\ & 2 (c_{i,j} + c_k + c_l) \\
\ge\ & 2 \big(p^{\rm eq}_i(\hat{\cal M}^{\rm eq}) + p^{\rm eq}_j(\hat{\cal M}^{\rm eq}) + p^{\rm eq}_k(\hat{\cal M}^{\rm eq}) + p^{\rm eq}_l(\hat{\cal M}^{\rm eq}) \big) \label{eq:eg-er-pf}
\end{align}
	
When $i = k$ (or $j = l$), Eqn.~(\ref{eq:eg-er-pf}) can also be proven straightforwardly, by omitting $k$ (or $l$, respectively).

Since equal cost-sharing mechanism is budget-balanced, $c({\cal M}) = \sum_{i \in {\cal M}} p^{\rm eq}_i({\cal M})$, summing over $\hat{\cal M}^{\rm eq} \backslash {\cal M}^\ast$ can obtain 
\begin{equation}
3 \cdot c({\cal M}^\ast) \ge 2 \cdot c(\hat{\cal M}^{\rm eq})
\end{equation}
\end{IEEEproof}

\medskip

\begin{customthm}{2} \label{thm:ega-er}
For egalitarian cost-sharing mechanism, let $\hat{\cal M}^{\rm ega}$ be a stable ride-sharing assignment. We show that the social optimality ratio is upper bounded by $\frac{c(\hat{\cal M}^{\rm ega})}{c({\cal M}^\ast)} \le \frac{3}{2}$.
%\begin{equation}
%\frac{3}{2} \ge \frac{c(\hat{\cal M}^{\rm ega})}{c({\cal M}^\ast)}
%\end{equation}
\end{customthm}

\medskip

\begin{customthm}{3} \label{thm:pp-er}
For proportional cost-sharing mechanism, let $\hat{\cal M}^{\rm pp}$ be a stable ride-sharing assignment. We show that the social optimality ratio is upper bounded by $\frac{c(\hat{\cal M}^{\rm pp})}{c({\cal M}^\ast)} \le \frac{3}{2}$.
%\begin{equation}
%\frac{3}{2} \ge \frac{c(\hat{\cal M}^{\rm pp})}{c({\cal M}^\ast)}
%\end{equation}
\end{customthm}

\medskip

\begin{customthm}{4} \label{thm:sb-er}
For segment-based cost-sharing mechanism, let $\hat{\cal M}^{\rm sb}$ be a stable ride-sharing assignment.  The corresponding social optimality ratio is upper bounded by $\frac{c(\hat{\cal M}^{\rm sb})}{c({\cal M}^\ast)} \le \frac{3}{2}$.
%\begin{equation}
%\frac{3}{2} \ge \frac{c(\hat{\cal M}^{\rm sb})}{c({\cal M}^\ast)}
%\end{equation}
\end{customthm}

\medskip

The proofs of Theorems~\ref{thm:ega-er}-\ref{thm:sb-er} can be found in the appendix.

{\bf Remark:} Theorems~\ref{thm:eq-er}-\ref{thm:sb-er} show that the social optimality ratios under equal, egalitarian, proportional and segment-based cost-sharing mechanisms are at most $\frac{3}{2}$ in any instances of any number of commuters, which is a small constant. Therefore, these fair cost-sharing mechanisms can achieve high social optimality. Note that in practice, the social optimality ratios are much smaller than the theoretical bounds, and hence, can achieve even higher social optimality.

\section{Stable Matching Algorithm}  \label{sec:matching}

To complement our analysis of stable matching on ride-sharing, we present a modified stable matching algorithm for finding a stable ride-sharing assignment. This algorithm is based on the classical Gale-Shapley algorithm for stable marriage problem and Irving's algorithm for stable roommates problem \cite{stablematchbook}. Here, we extend the classical algorithms to allow the possibility of standalone rides (namely, with no ride-sharing partner). The modified stable matching algorithm will be used in Sec.~\ref{sec:empirical} for an empirical data analysis on taxi sharing in New York City.

We first define some notations. For each $i \in {\cal N}$, define a preferential order ($ \succ_i$) over all possible options of ride-sharing pairs with $i$ (including standalone ride). For example, ``$(i, j) \succ_i (i, k) \succ_i (i, i) \succ_i ...$'' means that $(i, j)$ is the most preferred by $i$, then followed by $(i, k)$ and standalone ride $(i, i)$, and so on. Each commuter's preferential order is formulated according to the ranking order of individual payments of a given cost-sharing mechanism (e.g., $p_i(r^{\min}_{i, j}) < p_i(r^{\min}_{i, k}) < p_i(r^{\rm self}_{i}) < ... $). For equal payments, we will enforce deterministic tie-breaking in a consistent manner across all commuters. Note that the preferential orders also include the option of standalone ride $r^{\rm self}_{i}$. Each commuter will remove the options of ride-sharing below the standalone ride, as they will not be selected at the end.

The stable matching algorithm is consisted of several rounds. Initially, all commuters are set to be {\em unsuspended}, which allows them to propose to any partners. In each round, first each unsuspended commuter $i$ will propose to a ride-sharing partner he prefers most in his preferential order to whom $i$ has not proposed in the previous round, and is better than the partner with whom $i$ is currently provisionally matched, if any. Note that for a standalone ride, $i$ will propose to himself. Next, each commuter $j$ will collect a number of proposals at each round. He will select the most preferred one from the received proposals. If commuter $j$ is currently provisionally matched with, say $l$, and $j$ prefers another new proposer $i$ to $l$, then $(j, l)$ will be unmatched and $(i, j)$ will be provisionally matched, instead. This process will continue until every commuter is provisionally matched with another ride-sharing partner. 

The pseudo-code of stable matching algorithm is described in {\sc StableMatching}.

\begin{algorithm}[htp!] 
\caption{ {\sc StableMatching}} \label{alg:stableMatching}
\begin{algorithmic}[1]
\Require Preferential orders $(\succ_i)_{i \in {\cal N}}$
\Ensure Ride-sharing assignment ${\cal M}$
\State Initialize $\forall i \in {\cal N}$ to be {\sc unsuspended}
\While{$\exists$ {\sc unsuspended} $i$ who still has a potential ride-sharing partner to propose} 
	\State  $j\leftarrow$ first potential partner on $i$'s preferential order to 
	\Statex \quad \ \ \qquad whom $i$ has not yet proposed 
	\State Suppose $(i, k)$ {\sc provisionally matched}
\If{$i$ prefers $j$ to $k$ (i.e., $(i, j) \succ_i (i, k)$)}
	\State $i$ proposes to $j$
	\If{$j$ is not {\sc provisionally matched}}
		\State $(i, j)$ is {\sc provisionally matched}
		\State Set $i$ to be {\sc suspended}
	\Else
		\If{$\exists$ $(j, l)$ {\sc provisionally matched}}
			\If{$j$ prefers $i$ to $l$ (i.e., $(i, j) \succ_j (j, l)$)}
				\State Set $l$ to  {\sc unsuspended}
				\State $(i, j)$ is {\sc provisionally matched}
				\State $(j, l)$ is {\sc unmatched}
			\Else
				\State $(j, l)$ still {\sc provisionally matched}

			\EndIf
		\EndIf
	\EndIf
\EndIf
\EndWhile	
\State Set ${\cal M}$ to include all {\sc provisionally matched} pairs
\end{algorithmic}
\end{algorithm}

Next, we define a {\em cyclic preference} as a sequence of commuters $(i_1,..., i_s)$, such that
\begin{equation}
(i_1, i_s) \succ_{i_1} (i_1, i_2), \
(i_1, i_2) \succ_{i_2} (i_2, i_3), \
 \ldots, \  
(i_{s-1}, i_s) \succ_{i_s} (i_1, i_s) 
\end{equation}

\begin{customthm}{5} \label{thm:no-cyclic}
If there exists no cyclic preference, then {\sc StableMatching} will converge to a stable matching outcome.
\end{customthm}

\medskip
The full proof of existence of stable matching outcome under the condition of no cyclic preference can be found in \cite{CE17sharing}. The basic idea is that the absence of cyclic preference rules out the possibility of oscillation, where commuters keep switching matched pairs without termination.
Note that there is no cyclic preference under equal, egalitarian and proportional cost-sharing mechanisms, as shown in \cite{CE17sharing}. Although segment-based cost-sharing mechanism may induce cyclic preference, this is uncommon in practice.

Although {\sc StableMatching} is an extension of the classical matching algorithm from Irving \cite{stablematchbook}, there are some subtle differences in our algorithm. First, our algorithm allows any commuter to be unmatched (and hence taking a standalone ride by himself). Second, Theorem~\ref{thm:no-cyclic} shows that the common fair cost-sharing mechanisms will induce no cyclic preferences. Hence, there is no need to deal with cyclic preferences by considering odd rotations in the original Irving's algorithm.

\section{Empirical Data Analysis of Taxi Sharing} \label{sec:empirical}

To corroborate our theoretical results of social optimality ratios in Sec.~\ref{sec:empirical}, we present an empirical big data analysis on practical taxi sharing in NYC. We compare various properties of different fair cost-sharing mechanisms used for taxi sharing in an empirical study based on NYC taxi trip dataset \cite{nycdb}. We provide useful insights on effective cost-sharing mechanisms for ride-sharing in practice. 

In our data analysis, we used the taxi trip dataset of NYC (NYC) of 2013 \cite{nycdb}. The dataset contains over 450K taxi trips per day with the average distance per trip is around 4.2 km. Each data record of a trip includes the information of Taxi ID, trip distance and duration times of pick-ups and drop-offs of commuters as well as the GPS locations of pick-ups and drop-offs of commuters.

\begin{figure}[t!]
\centering
\begin{subfigure}[b]{0.5\textwidth}
\includegraphics[width=1\textwidth]{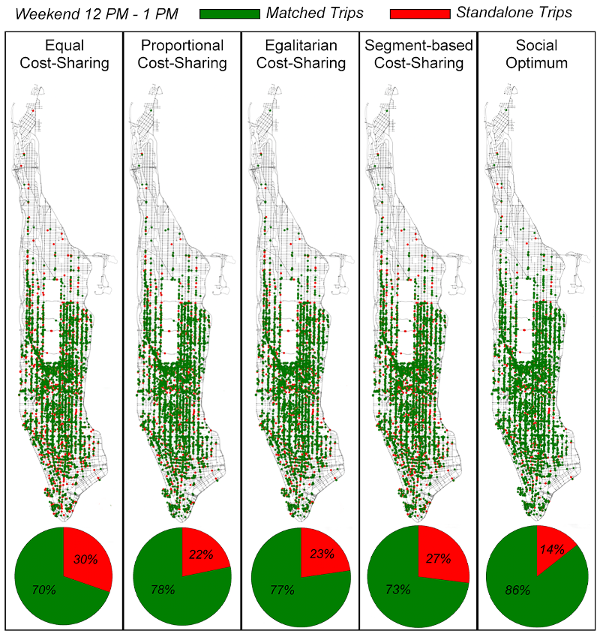}
\caption{Density maps of pick-up locations. The green dots indicate the pick-up locations of matched commuters, whereas the red dots indicate those of unmatched commuters.}
\label{fig:heatmap}
\bigskip
\end{subfigure} 
\begin{subfigure}[b]{0.55\textwidth}\hspace{-16pt}
\includegraphics[width=1.03\textwidth]{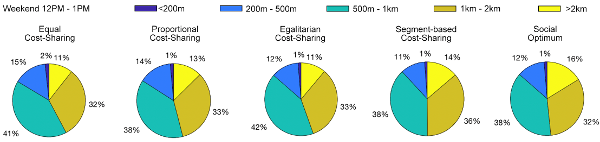}
\caption{Distributions of total separation distance in pick-up locations between \\ pairs of matched commuters.}
\label{fig:matchpiecharts}
\end{subfigure}
\smallskip
\caption{(a) Density maps of the pick-up locations, and (b) distributions of total separation distance between matched commuters based on different cost-sharing mechanisms and social optimum.} \label{fig:heatmappiecharts}
\end{figure}

\subsection{Density Maps of Matched Commuters}

{\bf Setting:} 
We first present a case study of the effectiveness of equal, proportional, egalitarian and segment-based cost-sharing mechanisms, as compared to the social optimal outcome. We consider the taxi trips during 12pm-1pm on 23th Feb 2013 (weekend) with over 5000 taxi trips. We employed the modified stable matching algorithm in Sec.~\ref{sec:matching} to find stable ride-sharing assignments. We consider a pair of commuters who can be potentially matched, if their pick-up times in the NYC dataset are within 3 minutes with each other. We examine the properties of matched and unmatched commuters in stable matching outcomes in Fig.~\ref{fig:heatmap}. We compare the outcomes with the social optimal outcome, which minimizes the total cost of all commuters. We also examine the distributions of total separation distance in pick-up locations between pairs of matched commuters in Fig.~\ref{fig:matchpiecharts}.

{\bf Observations:} 
To visualize the outcomes of ride-sharing, we plot five density maps of pick-up locations of commuters in Fig.~\ref{fig:heatmap}. Not all commuters can be matched for ride-sharing. Some of them have standalone rides. The green dots indicate the pick-up locations of matched commuters, whereas the red dots indicate those of unmatched commuters. 

We observe that the density maps of different cost-sharing mechanisms are rather similar, and most of the pick-up locations of matched commuters are located in similar places. However, the portions of matched commuters are different. Over 70\% of commuters can be matched based on any of the four cost-sharing mechanisms. However, equal and segment-based cost-sharing mechanisms can match a fewer number of commuters (27\%-30\%) than proportional and egalitarian cost-sharing mechanisms (22\%-23\%), which are fewer than the social optimal outcome (14\%). Since equal cost-sharing mechanism can induce negative utility, it may discourage sharing between commuters, and lead to lower percentage of matched pairs. On the other hand, social optimal matching ignores stability in matching, and hence can match the maximum number of pairs of commuters.

%We also plot the distributions of total separation distance in pick-up locations between pairs of matched commuters in Fig.~\ref{fig:matchpiecharts}. We observe that most matched commuters are separated by less than 2km in their pick-up locations. We also notice that over 40\% of matched commuters are separated by over 500m in their pick-up locations and segment-based cost-sharing mechanism induce the highest percentage of  separated matched commuters (50\% over 500m). 

\begin{figure*}[ht]
\centering
\begin{subfigure}[b]{0.19\textwidth}
\includegraphics[width=1\textwidth]{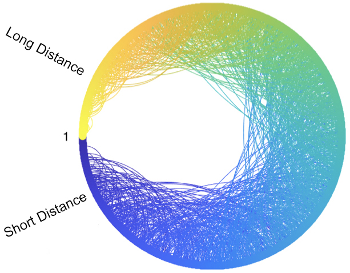}
\caption{Equal.}
\label{fig:coal_eq}
\end{subfigure} 
\begin{subfigure}[b]{0.19\textwidth}
\includegraphics[width=1\textwidth]{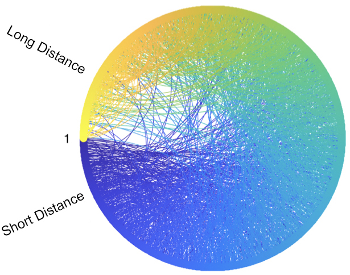}
\caption{Proportional.}
\label{fig:coal_pp}
\end{subfigure} 
\begin{subfigure}[b]{0.19\textwidth}
\includegraphics[width=1\textwidth]{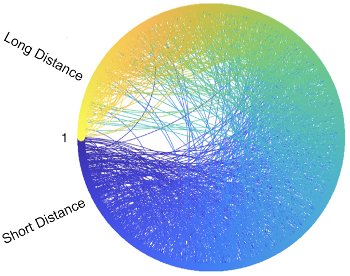}
\caption{Egalitarian.}
\label{fig:coal_ega}
\end{subfigure} 
\begin{subfigure}[b]{0.19\textwidth}
\includegraphics[width=1\textwidth]{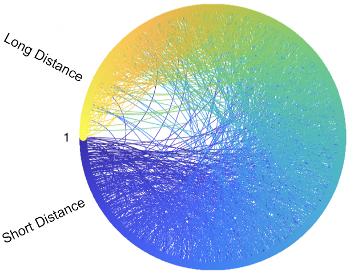}
\caption{Segment-based.}
\label{fig:coal_sb}
\end{subfigure} 
\begin{subfigure}[b]{0.19\textwidth}
\includegraphics[width=1\textwidth]{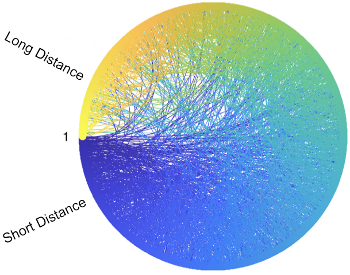}
\caption{Social optimum.}
\label{fig:coal_se}
\end{subfigure} 
\caption{Stable matching structures based on different cost-sharing mechanisms and social optimal outcome. Each commuter is represented by a point on the perimeter of a circle, following the order of distances of standalone rides. An edge is drawn between a pair of matched commuters.} \label{fig:coal}
\end{figure*}

\begin{figure*}[ht]
\centering
\begin{subfigure}[b]{0.49\textwidth}
\includegraphics[width=1\textwidth]{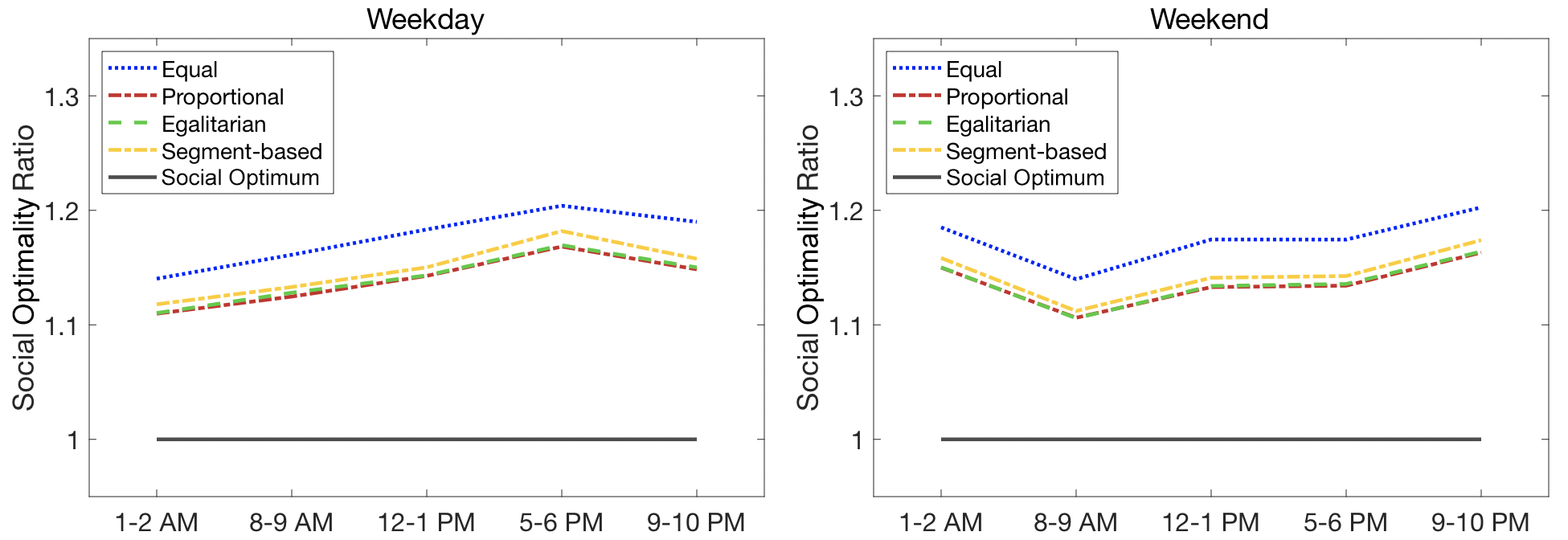}
\caption{Social optimality ratios.}
\label{fig:seratio}
\end{subfigure}
\begin{subfigure}[b]{0.49\textwidth}
\includegraphics[width=1\textwidth]{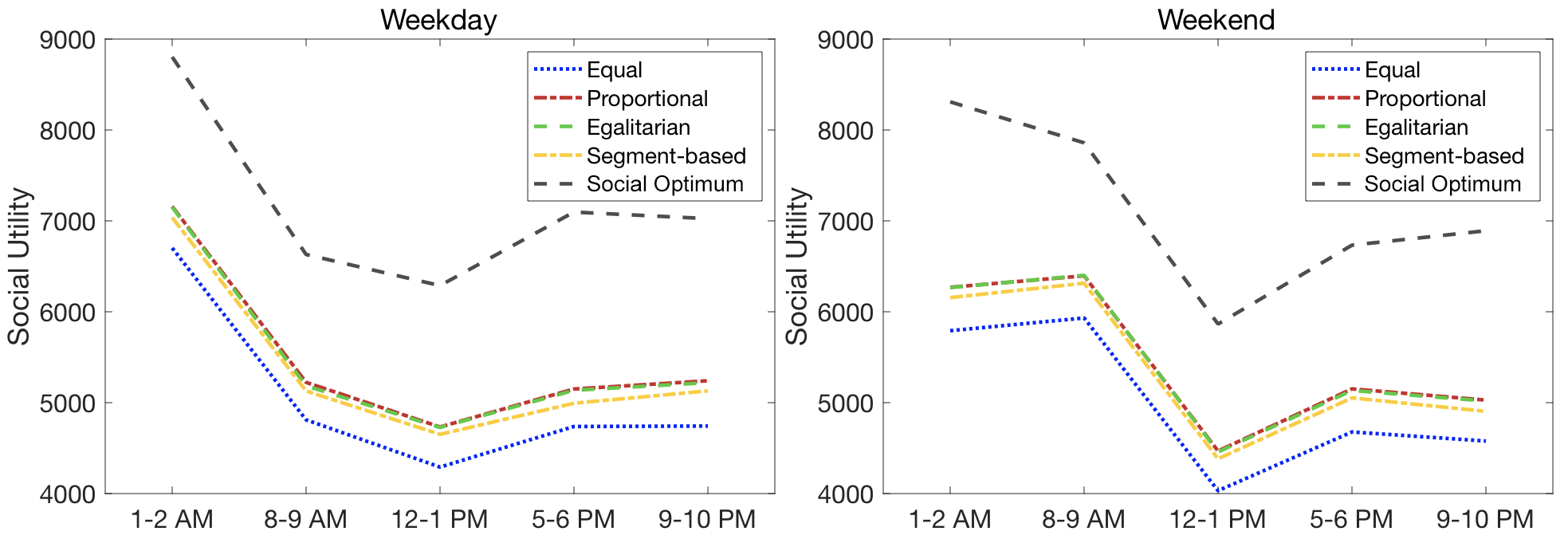}
\caption{Social utilities.}
\label{fig:utility}
\end{subfigure} 
\caption{(a) Social optimality ratios and (a) social utilities based on different cost-sharing mechanisms.} \label{fig:utilityseratio}
\end{figure*}

\subsection{Stable Matching Structures}

{\bf Setting:}  We next study the structures of stable matching outcomes under different cost-sharing mechanisms. We first sorted the matched commuters according to the distances of their standalone rides. In Fig.~\ref{fig:coal}, each commuter is represented by a point on the perimeter of a circle, following the order of distances of standalone rides. An edge is drawn between a pair of matched commuters. We also color the commuters with long standalone rides by yellow, and the commuters with short standalone rides by blue.

{\bf Observations:}  In Fig.~\ref{fig:coal}, we visualize the stable matching structures. We observe that different cost-sharing mechanisms induce different stable matching structures. For equal cost-sharing mechanism, commuters with long standalone rides are more likely to be matched among themselves, and so are those with short standalone rides. On the other hand, for proportional, egalitarian and segment-based cost-sharing mechanisms, commuters with long standalone rides are more likely to be matched with those of short standalone rides. Hence, proportional, egalitarian and segment-based cost-sharing mechanisms can bolster diversity in stable matching with heterogeneous commuters of different pick-up and drop-off locations.

\subsection{Social Optimality}

{\bf Setting:}  We next examine the social optimality of different cost-sharing mechanisms for practical taxi sharing. The study is based on the data of January 4, 2013 (i.e., Weekday) and February 23, 2013 (i.e., Weekend) in the NYC taxi trip dataset. We selected five representative one-hour periods (1-2 am, 8-9am, 12-1pm, 5-6pm, 9-10pm) on each day. In Fig.~\ref{fig:seratio}, we compare the social optimality ratios of the social costs over the costs of social optimal outcomes in the respective periods. In Fig.~\ref{fig:utility}, we also compare the social utilities of all matched commuters (which is the total savings from standalone rides of all commuters: $\sum_i u_i = \sum_i c_i - p_i$).

{\bf Observations:}  We observe that the empirical social optimality ratios of four cost-sharing mechanisms ($\le$ 1.2) are well below the theoretical upper bound $\frac{3}{2}$ in Sec.~\ref{sec:empirical}. Indeed, the social optimality ratios are much in practice smaller than the theoretical bounds, and hence, can achieve higher social optimality. In particular, equal cost-sharing mechanism induce less social utilities and larger social optimality ratios, whereas the other three cost-sharing mechanisms produce comparable results. In terms of social utilities, we observe that all cost-sharing mechanisms can achieve higher social utilities, when the social optimal outcome has a higher social utility.

\begin{figure*}[ht]
\centering
\begin{subfigure}[b]{0.49\textwidth}
\includegraphics[width=1\textwidth]{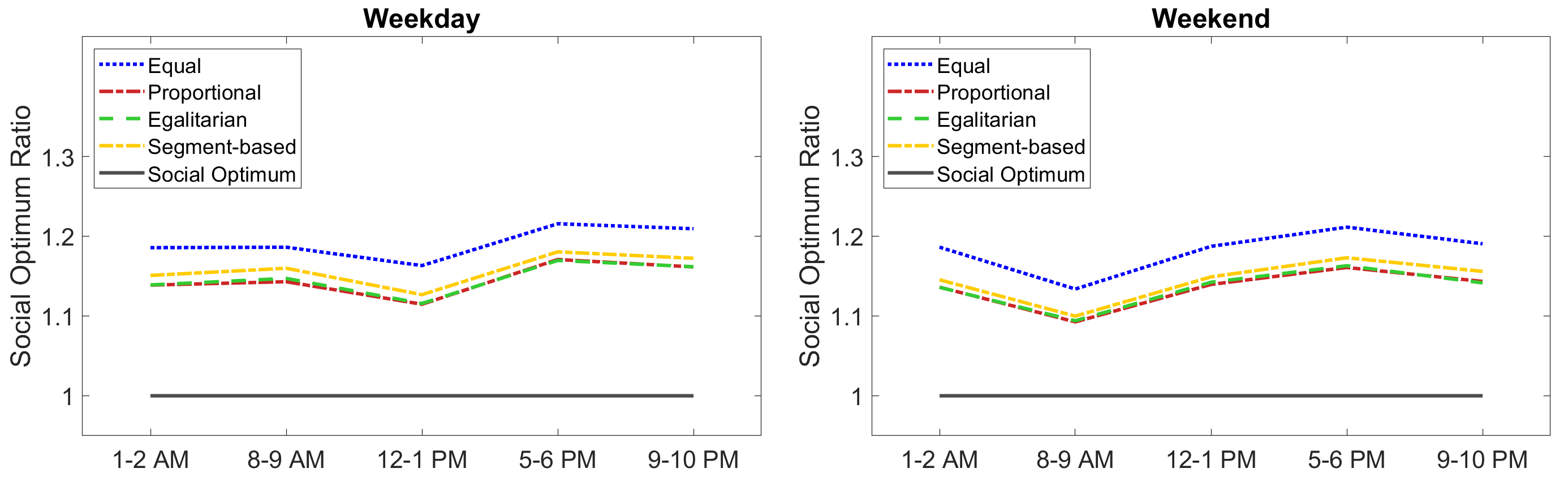}
\caption{Social optimality ratios (with 40\% subsampled data).}
\label{fig:seratio40}
\end{subfigure}
\begin{subfigure}[b]{0.49\textwidth}
\includegraphics[width=1\textwidth]{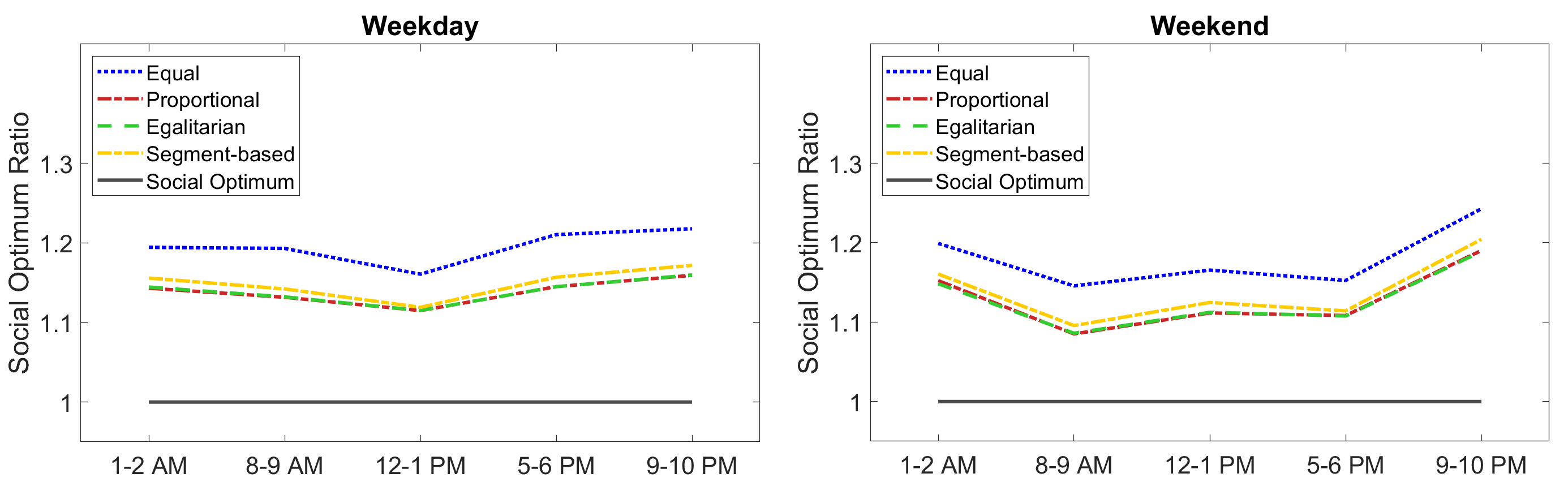}
\caption{Social optimality ratios (with 20\% subsampled data).}
\label{fig:seratio20}
\end{subfigure} 
\caption{Social optimality ratios with subsampled data.} \label{fig:utilityseratio}
\end{figure*}

We also carried out our evaluation on randomly sub-sampled dataset with 40\% and 20\% data to validate the social optimality ratios in Figs.~\ref{fig:seratio40}-\ref{fig:seratio20}. All these results show very similar trends among different cost-sharing mechanisms, despite different data samples, which are also well below the theoretical upper bound $\frac{3}{2}$. These results provide the confidence in our mechanisms as a viable way for decentralized ride-sharing.

\begin{figure*}[ht]
\centering
\begin{subfigure}[b]{0.49\textwidth}
\includegraphics[width=1.\textwidth]{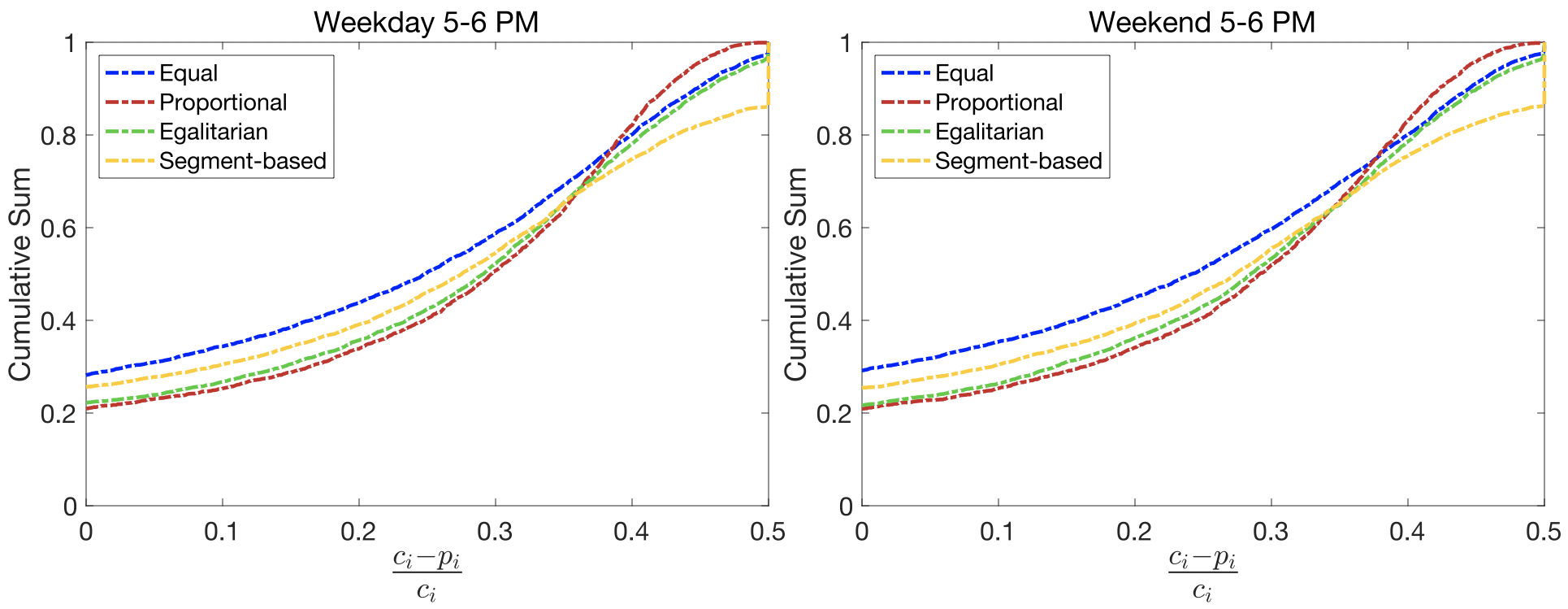}
\caption{Normalized utilities.}
\label{fig:savingratio}
\end{subfigure} 
\begin{subfigure}[b]{0.49\textwidth}
\includegraphics[width=1\textwidth]{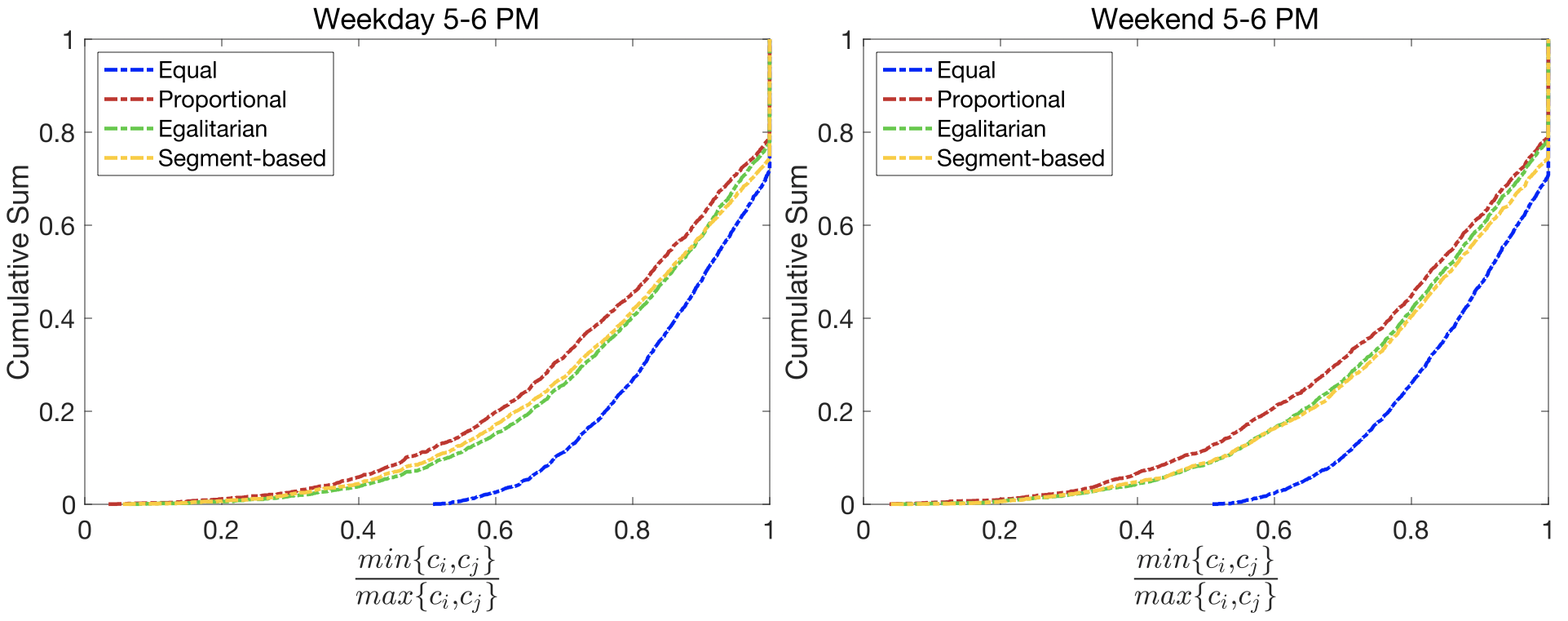}
\caption{Standalone cost ratios.}
\label{fig:diffratio}
\end{subfigure}
\caption{Cumulative distributions of (a) normalized utilities ($\frac{c_i - p_i}{c_i}$) and (b) standalone cost ratios $\frac{\min\{c_i, c_j\}}{\max\{c_i, c_j\}}$ under different cost-sharing mechanisms.} \label{fig:ratios}
\end{figure*}

\subsection{Distributions of Normalized Utilities}

{\bf Setting:}  
To shed light on the individual benefits of matched commuters, we define the normalized utility of commuter $i$ by the ratio $\frac{c_i - p_i}{c_i}$, which is the normalized utility over the standalone ride cost. In Fig.~\ref{fig:savingratio}, we compare the cumulative distributions of normalized utilities among the matched commuters under different cost-sharing mechanisms. 

{\bf Observations:} 
We observe that the cumulative distributions of normalized utilities under different cost-sharing mechanism on weekday and weekend are similar. Comparing the outcomes of different cost-sharing mechanisms, more commuters have lower normalized utilities under equal cost-sharing mechanism. The number of commuters who have the highest normalized utilities (0.5) under proportional mechanism is the largest among the four cost-sharing mechanisms, which is approximately 15\% of total commuters. Thus, proportional cost-sharing mechanism can benefit commuters with higher normalized savings. Furthermore, proportional and egalitarian cost-sharing mechanisms induce similar cumulative distributions of normalized utilities of less than 0.35, whereas proportional cost-sharing mechanism induces more commuters having the normalized utilities around 0.4 than egalitarian cost-sharing mechanism.

\subsection{Distributions of Standalone Cost Ratios}

{\bf Setting:}  
Next, we study the similarity between commuters in a matched pair. We define the standalone cost ratio of a matched pair of commuters $(i, j)$  by $\frac{\min\{c_i, c_j\}}{\max\{c_i, c_j\}}$, which is the maximum over different standalone ride costs for a pair $(i, j)$. If $\frac{\min\{c_i, c_j\}}{\max\{c_i, c_j\}} \to 1$, the matched pair are relatively similar. In Fig.~\ref{fig:diffratio}, we compare the cumulative distributions of standalone cost ratios among the matched commuters under different cost-sharing mechanisms. 

{\bf Observations:} The cumulative distributions of standalone cost ratios under different cost-sharing mechanism on weekday and weekend commuters are similar. Overall, commuters are more likely matched with those of similar standalone costs since more commuters have high standalone cost ratios ($>$ 0.8) in any one of the cost-sharing mechanisms. It’s also worth to notice that approximately 70\% of commuters have high standalone cost ratios ($>$ 0.8) under equal cost-sharing mechanism, whose percentage is more than that the other three cost-sharing mechanisms. Egalitarian and segment-based cost-sharing mechanisms produce the similar cumulative distributions of standalone cost ratios, whereas proportional cost-sharing mechanism produces more commuters who have lower standalone cost ratios than the other cost-sharing mechanisms, which indicates that commuters are more likely matched with those of different standalone costs under proportional cost-sharing mechanism.

\begin{figure}[b!]
\includegraphics[width=0.5\textwidth]{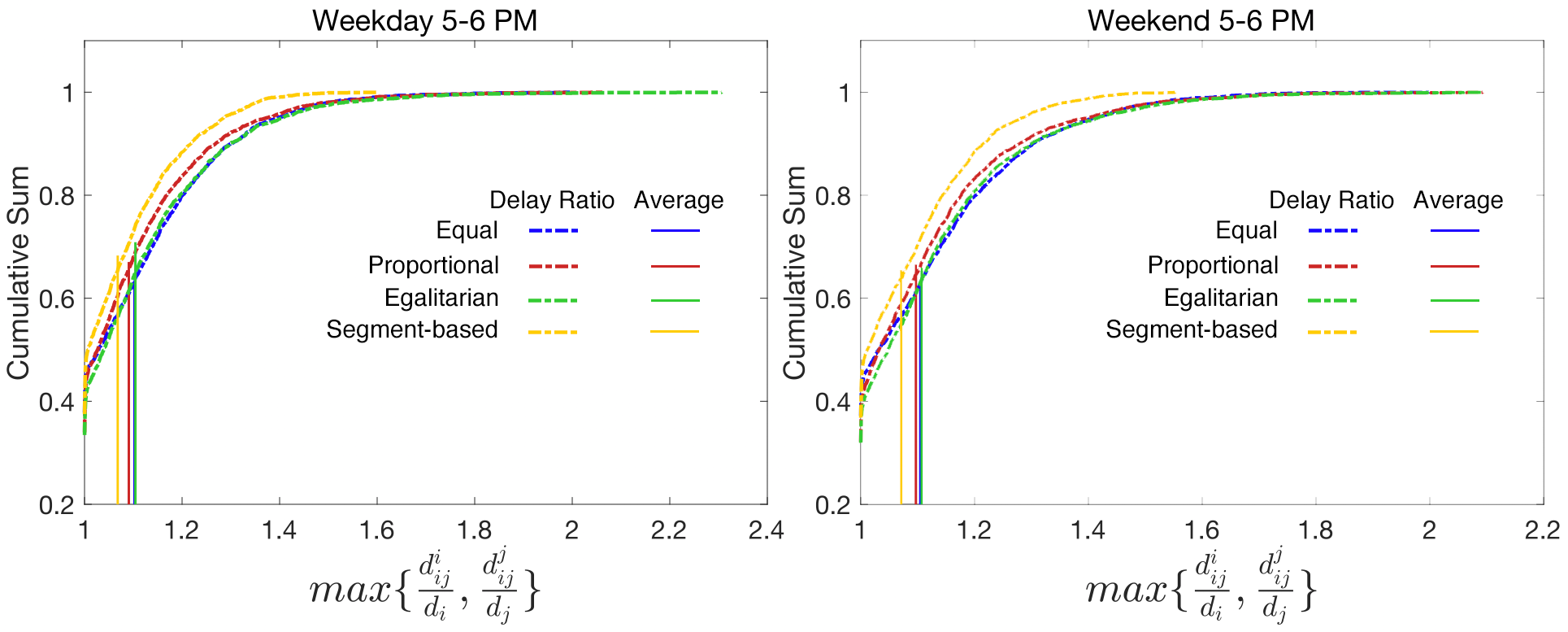}
\caption{Cumulative distributions of delay ratios ($\max\{ \frac{d^i_{i,j}}{d_i}, \frac{d^j_{i,j}}{d_j}\}$) under different cost-sharing mechanisms.} \label{fig:delayratio}
\end{figure}

\subsection{Distributions of Delay Ratios}

{\bf Setting:}  
Ride-sharing may incur additional delay to a commuter because of a detour to pick-up another commuter. Particularly, we consider the incurred delay in terms of additional geographical distance in ride-sharing, in the presence of similar traffic condition. We denote the geographical distances traveled of a matched pair of commuters $i$ and $j$ in a shared ride by $d^i_{i,j}$ and $d^i_{i,j}$ respectively. We denote the geographical distance by commuter $i$ in his standalone ride by $d_i$. Define the delay ratio of a matched pair of commuters $(i, j)$ by $\max\{ \frac{d^i_{i,j}}{d_i}, \frac{d^j_{i,j}}{d_j}\}$, which is a natural metric of delay in a shared ride. In Fig.~\ref{fig:delayratio}, we compare the cumulative distributions of delay ratios among the matched commuters under different cost-sharing mechanisms. 

{\bf Observations:} The cumulative distributions of delay ratios under different cost-sharing mechanisms on weekday and weekend are similar. We also notice that the average delay ratios are around 1.1, and over 90\% of matched commuters have delay ratios less than 1.4. Hence, small delays are incurred among the matched commuters. Segment-based cost-sharing mechanism induces more matched commuters having lower delay ratio ($<$ 1.1) than other three cost-sharing mechanisms. In addition, equal and egalitarian cost-sharing mechanisms produce similar cumulative distributions of delay ratios, whereas proportional cost-sharing mechanism produce more matched commuters who have slightly lower delay ratios than equal and egalitarian cost-sharing mechanisms.

\subsection{Insights and Ramifications}

Overall, we draw the following insights and ramifications based on our empirical study with NYC taxi trip dataset:
\begin{enumerate}

\item The four fair (i.e., equal, egalitarian, proportional and segment-based) cost-sharing mechanisms can enable effective decentralized ride-sharing arrangement in practice by achieving high social optimality, such that the induced social costs are as comparably low as the social optimal arrangement by a centralized planner whose objective is to minimize the total cost.

\item Egalitarian and proportional cost-sharing mechanisms can induce more matched commuters than equal and segment-based cost-sharing mechanisms.

\item Egalitarian, proportional and segment-based cost-sharing mechanisms can bolster diversity among the matched commuters with heterogeneous locations.

\item Proportional cost-sharing mechanism can benefit commuters with higher normalized savings.

\item All four cost-sharing mechanisms incur small delays among the matched commuters.

\end{enumerate}

\section{Conclusion}\label{sec:concl}

Ride-sharing is a popular paradigm, requiring decentralized decision-making processes among commuters. This paper offers a thorough study of decentralized ride-sharing arrangements based on the principle of fair cost-sharing of transportation costs. We define several fair cost-sharing mechanisms, including equal, egalitarian, proportional and segment-based cost-sharing mechanisms. We compare the stable matching outcomes induced by these cost-sharing mechanisms with a social optimal outcome by deriving the theoretical bounds of social optimality ratios. Our results show that these fair cost-sharing mechanisms can achieve high social optimality. We corroborate our results with an empirical study of taxi sharing under fair cost-sharing mechanisms with New York City taxi trip dataset. In particular, our empirical study that egalitarian and proportional cost-sharing mechanisms can approach a social optimal outcome, although all equal, egalitarian, proportional and segment-based cost-sharing mechanisms share the same theoretical bound of social optimality ratio. 

Apart from the study of social optimality ratio, there are other potential areas to explore in future work. %First, our theoretical analysis does not consider the shareability network structure of feasible coalitions. Some network structures may favor a particular type of cost-sharing mechanism. 
This work so far considers transportation cost as the primary factor of coalition formation. Another possible factor is delay, which can lead to different coalition structures. %Third, we consider stable matching for ride-sharing with a pair of commuters. 
There are other possibilities, for example, ride-sharing with more than two commuters in each group, and between drivers and passengers, who have asymmetric roles. Also, we will consider uncertain information (e.g., traffic information) and unknown arrivals of future commuters in an online fashion.

\appendix

\begin{customthm}{2} 
For egalitarian cost-sharing mechanism, let $\hat{\cal M}^{\rm ega}$ be a stable ride-sharing assignment. We show that the social optimality ratio is upper bounded by $\frac{c(\hat{\cal M}^{\rm ega})}{c({\cal M}^\ast)} \le \frac{3}{2}$.
\end{customthm}

\begin{IEEEproof}
Similar to the proof of Theorem~\ref{thm:eq-er}, suppose $(i, j) \in \hat{\cal M}^{\rm ega} \backslash {\cal M}^\ast$. Then there must exist $(i, k)$ and $(j, l)$, such that $(i, k), (j, l) \in {\cal M}^\ast \backslash \hat{\cal M}^{\rm ega}$. We also obtain
\begin{align}
 & p^{\rm ega}_i(\hat{\cal M}^{\rm ega}) + p^{\rm ega}_j(\hat{\cal M}^{\rm ega}) + p^{\rm ega}_k(\hat{\cal M}^{\rm ega}) + p^{\rm ega}_l(\hat{\cal M}^{\rm ega}) \notag \\
 \le & \ c_{i, j} + c_{k} + c_{l}
\end{align}

Since $\hat{\cal M}^{\rm ega}$ is a stable ride-sharing assignment, we obtain
\begin{align}
 \tfrac{c_{i,j} + c_i - c_j}{2} = p^{\rm ega}_i(\hat{\cal M}^{\rm ega})  \le \ & p^{\rm ega}_i({\cal M}^\ast) = \tfrac{c_{i,k} + c_i - c_k}{2} \notag \\ 
\Rightarrow \quad c_{i,j} + c_k   \le \ & c_{i,k} + c_j \\
 \tfrac{c_{i,j} + c_i - c_j}{2} = p^{\rm ega}_j(\hat{\cal M}^{\rm ega})  \le \ & p^{\rm ega}_j({\cal M}^\ast) = \tfrac{c_{j,l} + c_i - c_l}{2} \notag \\ 
\Rightarrow \quad c_{i,j} + c_l  \le \ & c_{j,l} + c_j 
\end{align}

Together, by noting that $c_{i, k} \ge \max\{c_i, c_k\}$ and $c_{j, l} \ge \max\{c_j, c_l\}$, we obtain
\begin{align}
&  3  \big(p^{\rm ega}_i({\cal M}^\ast) + p^{\rm ega}_j({\cal M}^\ast) + p^{\rm ega}_k({\cal M}^\ast) + p^{\rm ega}_l({\cal M}^\ast) \big) \notag \\
=\ & 3 (c_{i, k} + c_{j, l})\\
\ge\ & c_{i, k} + c_{j, l} + c_i + c_j + c_k + c_l \\
\ge\ &  2 (c_{i,j} + c_k + c_l) \\
\ge\ & 2 \big(p^{\rm ega}_i(\hat{\cal M}^{\rm ega}) + p^{\rm ega}_j(\hat{\cal M}^{\rm ega}) + p^{\rm ega}_k(\hat{\cal M}^{\rm ega}) + p^{\rm ega}_l(\hat{\cal M}^{\rm ega}) \big)
\end{align}

Therefore, similar to Theorem~\ref{thm:eq-er}, it completes the proof.
\end{IEEEproof}

\smallskip

\begin{customthm}{3}
For proportional cost-sharing mechanism, let $\hat{\cal M}^{\rm pp}$ be a stable ride-sharing assignment. We show that the social optimality ratio is upper bounded by $\frac{c(\hat{\cal M}^{\rm pp})}{c({\cal M}^\ast)} \le \frac{3}{2}$.
\end{customthm}

\begin{IEEEproof}
Similar to the proof of Theorem~\ref{thm:eq-er}, suppose $(i, j) \in \hat{\cal M}^{\rm pp} \backslash {\cal M}^\ast$. Then there must exist $(i, k)$ and $(j, l)$, such that $(i, k), (j, l) \in {\cal M}^\ast \backslash \hat{\cal M}^{\rm pp}$. We also obtain
\begin{equation}
 p^{\rm pp}_i(\hat{\cal M}^{\rm pp}) + p^{\rm pp}_j(\hat{\cal M}^{\rm pp}) + p^{\rm pp}_k(\hat{\cal M}^{\rm pp}) + p^{\rm pp}_l(\hat{\cal M}^{\rm pp}) \le c_{i, j} + c_{k} + c_{l}
\end{equation}

Since $\hat{\cal M}^{\rm pp}$ is a stable ride-sharing assignment, we obtain
\begin{align}
 \tfrac{c_i \cdot c_{i,j}}{c_i + c_j} = p^{\rm pp}_i(\hat{\cal M}^{\rm pp})  \le \ & p^{\rm pp}_i({\cal M}^\ast) = \tfrac{c_i \cdot c_{i,k}}{c_i + c_k} \notag \\ 
\Rightarrow \quad \tfrac{c_{i,j}(c_i + c_k)}{c_i + c_j}   \le \ & c_{i,k} \\
 \tfrac{c_j \cdot c_{i,j}}{c_i + c_j} = p^{\rm pp}_j(\hat{\cal M}^{\rm pp})  \le \ & p^{\rm pp}_j({\cal M}^\ast) = \tfrac{c_j \cdot c_{j,l}}{c_j + c_l} \notag \\ 
\Rightarrow \quad \tfrac{c_{i,j}(c_j + c_l)}{c_i + c_j}  \le \ & c_{j,l}
\end{align}

Together, by noting that $c_{i, k} \ge \max\{c_i, c_k\}$ and $c_{j, l} \ge \max\{c_j, c_l\}$, we obtain
\begin{align}
&  3  \big(p^{\rm pp}_i({\cal M}^\ast) + p^{\rm pp}_j({\cal M}^\ast) + p^{\rm pp}_k({\cal M}^\ast) + p^{\rm pp}_l({\cal M}^\ast) \big) \notag \\
=\ & 3 (c_{i, k} + c_{j, l})\\
\ge\ & 2\Big( \frac{c_{i,j}(c_i + c_k)}{c_i + c_j} + \frac{c_{i,j}(c_j + c_l)}{c_i + c_j} \Big) + c_{i, k} + c_{j, l} \\
=\ & \tfrac{1}{c_i + c_j} (2 c_{i,j} c_i + 2 c_{i,j} c_k + 2 c_{i,j} c_j + 2 c_{i,j} c_l \\
& \qquad \qquad + c_{i,k} c_i + c_{i,k} c_j + c_{j,l} c_i + c_{j,l} c_j) \\
\ge\ & \tfrac{1}{c_i + c_j} (2 c_{i,j} c_i + 2 c_{i,j} c_k + 2 c_{i,j} c_j + 2 c_{i,j} c_l \\
& \qquad \qquad + c_{k} c_i + c_{k} c_j + c_{l} c_i + c_{l} c_j) \\
\ge\ & \tfrac{1}{c_i + c_j} (2 c_{i,j} c_i + c_{i} c_k + c_{j} c_k + 2 c_{i,j} c_j + c_{i} c_l + c_{j} c_l \\
& \qquad \qquad + c_{k} c_i + c_{k} c_j + c_{l} c_i + c_{l} c_j) \\
\ge\ &  2 (c_{i,j} + c_k + c_l) \\
\ge\ & 2  \big(p^{\rm pp}_i(\hat{\cal M}^{\rm pp}) + p^{\rm pp}_j(\hat{\cal M}^{\rm pp}) + p^{\rm pp}_k(\hat{\cal M}^{\rm pp}) + p^{\rm pp}_l(\hat{\cal M}^{\rm pp}) \big)
\end{align}

Therefore, similar to Theorem~\ref{thm:eq-er}, it completes the proof.
\end{IEEEproof}

\smallskip

\begin{customthm}{4}
For segment-based cost-sharing mechanism, let $\hat{\cal M}^{\rm sb}$ be a stable ride-sharing assignment. We show that the social optimality ratio is upper bounded by $\frac{c(\hat{\cal M}^{\rm sb})}{c({\cal M}^\ast)} \le \frac{3}{2}$.
\end{customthm}

\begin{IEEEproof}
Similar to the proof of Theorem~\ref{thm:eq-er}, suppose $(i, j) \in \hat{\cal M}^{\rm sb} \backslash {\cal M}^\ast$. Then there must exist $(i, k)$ and $(j, l)$, such that $(i, k), (j, l) \in {\cal M}^\ast \backslash \hat{\cal M}^{\rm sb}$. We also obtain
\begin{equation}
 p^{\rm sb}_i(\hat{\cal M}^{\rm sb}) + p^{\rm sb}_j(\hat{\cal M}^{\rm sb}) + p^{\rm sb}_k(\hat{\cal M}^{\rm sb}) + p^{\rm sb}_l(\hat{\cal M}^{\rm sb}) \le c_{i, j} + c_{k} + c_{l}
\end{equation}

Next, according to segment-based cost-sharing mechanism, each commuter $i$ contributes equally to the cost of the segments he participates in. Let $c_{i,k}(v^{\rm s}_k, v^{\rm d}_k)$ be the cost from the $k$'s source location to $k$'s destination location in the sharable ride for $(i,k)$. Note that $c_{i,k}(v^{\rm s}_k, v^{\rm d}_k) \ge c(r^{\rm self}_{k}) = c_k$. Otherwise, the commuter $k$ can always choose another lower cost standalone trip $r^{\rm self}_{k}$. 
We consider two cases:
\begin{itemize}

\item If $(i,k)$ share a hitchhiking ride, then
\begin{equation}
p^{\rm sb}_k({\cal M}^\ast) \ge \frac{c_{i,k}(v^{\rm s}_k, v^{\rm d}_k)}{2} \ge  \frac{c_k}{2}
\end{equation}

\item If $(i,k)$ share a combined ride, then 
\begin{align}
p^{\rm sb}_k({\cal M}^\ast) \ge & \min\Big\{c_{i,k}(v^{\rm s}_k, v^{\rm s}_i) + \tfrac{c_{i,k}(v^{\rm s}_i, v^{\rm d}_k) }{2}, \notag \\
  & \qquad \quad \tfrac{c_{i,k}(v^{\rm s}_k, v^{\rm d}_i) }{2} + c_{i,k}(v^{\rm d}_i, v^{\rm d}_k) \Big\}  \\
\ge & \tfrac{c_{i,k}(v^{\rm s}_k, v^{\rm d}_k)}{2} \ge  \tfrac{c_k}{2} 
\end{align}

\end{itemize}
Hence, we obtain $p^{\rm sb}_k({\cal M}^\ast) \ge \frac{c_k}{2}$ and similarly $p^{\rm sb}_l({\cal M}^\ast) \ge \frac{c_l}{2}$.

since $\hat{\cal M}^{\rm sb}$ is a stable ride-sharing assignment, we obtain
\begin{align}
 p^{\rm sb}_k(\hat{\cal M}^{\rm sb}) \le c_k, \quad p^{\rm sb}_l(\hat{\cal M}^{\rm sb}) \le c_l
\end{align}

Together, by noting that $c_{i, k} \ge \max\{c_i, c_k\}$ and $c_{j, l} \ge \max\{c_j, c_l\}$, we obtain
\begin{align}
&  3 \big(p^{\rm sb}_i({\cal M}^\ast) + p^{\rm sb}_j({\cal M}^\ast) + p^{\rm sb}_k({\cal M}^\ast) + p^{\rm sb}_l({\cal M}^\ast) \big) \notag \\
\ge\ & 2 \big(p^{\rm sb}_i({\cal M}^\ast) + p^{\rm sb}_j({\cal M}^\ast) + \tfrac{c_k}{2} + \tfrac{c_l}{2} \big)  + c_{i, k} + c_{j, l} \\
\ge\ & 2 (c_{i, j} + \tfrac{c_k}{2} + \tfrac{c_l}{2} )  + c_k + c_l \\
\ge\ & 2 \big(p^{\rm sb}_i(\hat{\cal M}^{\rm sb}) + p^{\rm sb}_j(\hat{\cal M}^{\rm sb}) + p^{\rm sb}_k(\hat{\cal M}^{\rm sb}) + p^{\rm sb}_l(\hat{\cal M}^{\rm sb}) \big)
\end{align}

Therefore, similar to Theorem~\ref{thm:eq-er}, it completes the proof.
\end{IEEEproof}

\bibliographystyle{ieeetr}
\bibliography{reference}

\begin{thebibliography}{10}

\bibitem{S14pnas}
P.~Santi, G.~Resta, M.~Szell, S.~Sobolevsky, S.~H. Strogatz, and C.~Ratti,
  ``Quantifying the benefits of vehicle pooling with shareability networks,''
  {\em PNAS}, vol.~111, pp.~13290--13294, Sep 2014.

\bibitem{stablematchbook}
D.~Gusfield and R.~W. Irving, {\em The Stable Marriage Problem: Structure and
  Algorithms}.
\newblock MIT Press, 1989.

\bibitem{nycdb}
{\relax NYC Taxi and Limousine Commission}, ``New york city taxi trip
  dataset,'' 2019.

\bibitem{NR2016ridesharing}
M.~Nourinejad and M.~J. Roorda, ``Agent based model for dynamic ridesharing,''
  {\em Transportation Research Part C}, vol.~64, pp.~117--132, 2016.

\bibitem{ZZ19matching}
H.~Zhang and J.~Zhao, ``Mobility sharing as a preference matching problem,''
  {\em IEEE Trans. Intell. Transp. Syst.}, vol.~20, July 2019.

\bibitem{P15matching}
D.~Pelzer, J.~Xiao, D.~Zehe, M.~Lees, A.~Knoll, and H.~Aydt, ``A
  partition-based match making algorithm for dynamic ridesharing,'' {\em IEEE
  Trans. Intell. Transp. Syst.}, vol.~16, pp.~2587--2598, May 2015.

\bibitem{W17matching}
X.~Wang, N.~Agatz, and A.~Erera, ``Stable matching for dynamic ride-sharing
  systems,'' {\em Transp. Sci.}, vol.~52, pp.~739--1034, Aug 2017.

\bibitem{agatz2020stableride}
N.~Agatz, A.~L. Erera, M.~W. Savelsbergh, and X.~Wang, ``Dynamic ride-sharing:
  a simulation study in metro atlanta,'' {\em Procedia Social and Behavioral
  Sciences 17 (2011) 532–550}, vol.~47, pp.~1--21, 2020.

\bibitem{wang2017matching}
X.~Wang, N.~Agatz, and A.~Erera, ``Stable matching for dynamic ride-sharing
  systems,'' {\em Transportation Science}, vol.~52, no.~4, 2017.

\bibitem{RC2019matching}
S.~Rasulkhani and J.~Y.~J. Chow, ``Route-cost-assignment with joint user and
  operator behavior as a many-to-one stable matching assignment game,'' {\em
  Transportation Research Part B}, vol.~124, pp.~60--81, 2019.

\bibitem{peng2020stableride}
Zixuan Peng, Wenxuan Shan, Peng Jia, Bin Yu, Yonglei Jiang, and
  Baozhen Yao, ``Stable ride-sharing matching for the commuters
  with payment design,'' {\em Transportation (2020) 47:1–21}, vol.~47,
  pp.~1--21, 2020.

\bibitem{A17pnas}
J.~Alonsomora, S.~Samaranayake, A.~Wallar, E.~Frazzoli, and D.~Rus, ``On-demand
  high-capacity ride-sharing via dynamic trip-vehicle assignment,'' {\em PNAS},
  vol.~114, pp.~462--467, 2017.

\bibitem{ADN09}
E.~Anshelevich, S.~Das, and Y.~Naamad, ``Anarchy, stability, and utopia:
  Creating better matchings,'' in {\em Algorithmic Game Theory {SAGT}},
  pp.~159--170, 2009.

\bibitem{AB12}
H.~Aziz and F.~Brandl, ``Existence of stability in hedonic coalition formation
  games,'' in {\em Proc. of the International Conference on Autonomous Agents
  and Multiagent Systems (AAMAS)}, 2012.

\bibitem{CE17sharing}
C.-K. Chau and K.~Elbassioni, ``Quantifying inefficiency of fair cost-sharing
  mechanisms for sharing economy,'' {\em IEEE Trans. Control of Network
  System}, vol.~5, pp.~1809--1818, Dec 2018.

\bibitem{CE20sharing}
S.~C.-K. Chau, K.~Elbassioni, and Y.~Zhou, ``Approximately socially-optimal
  decentralized coalition formation,'' tech. rep., Australian National
  University, 2019.

\bibitem{chau19p2penergy}
S.~C.-K. Chau, J.~Xu, W.~Bow, and K.~Elbassioni, ``Peer-to-peer energy sharing:
  Effective cost-sharing mechanisms and social efficiency,'' in {\em Proc. of
  ACM Intl. Conf. on Future Energy Systems (e-Energy)}, 2019.

\bibitem{cmtseng2019etaxi}
C.-M. Tseng, S.~C.-K. Chau, and X.~Liu, ``Improving viability of electric taxis
  by taxi service strategy optimization: A big data study of new york city,''
  {\em IEEE Trans. Intell. Transp. Syst.}, 2019.

\end{thebibliography}

\end{document}